\documentclass[aps,prb,twocolumn,superscriptaddress,showpacs]{revtex4-1}
\pdfoutput=1
\usepackage{textcomp}
\usepackage{bm}
\usepackage{amsmath}
\usepackage{amssymb}
\usepackage{graphicx}
\usepackage{color}
\usepackage{hyperref}
\hypersetup{colorlinks=true,linkcolor=red,citecolor=blue,urlcolor=blue}

\newcommand{\pd}{\partial}

\newcommand{\re}{\textrm{Re}}
\newcommand{\im}{\textrm{Im}}

\newcommand{\pv}{\text{P}}

\newcommand{\al}{\alpha}

\newcommand{\de}{\delta}

\newcommand{\la}{\lambda}
\newcommand{\om}{\omega}
\newcommand{\ka}{\kappa}

\newcommand{\Om}{\Omega}

\newcommand{\rra}{\rightarrow}

\makeatletter
\@ifundefined{textcolor}{}
{%
 \definecolor{BLACK}{gray}{0}
 \definecolor{WHITE}{gray}{1}
 \definecolor{RED}{rgb}{1,0,0}
 \definecolor{GREEN}{rgb}{0,1,0}
 \definecolor{BLUE}{rgb}{0,0,1}
 \definecolor{CYAN}{cmyk}{1,0,0,0}
 \definecolor{MAGENTA}{cmyk}{0,1,0,0}
 \definecolor{YELLOW}{cmyk}{0,0,1,0}
}

\makeatother

\begin{document}

\title{Thermodynamics of quantum crystalline membranes}

\author{B. Amorim}
\email[Electronic address: ]{amorim.bac@icmm.csic.es}
\affiliation{Instituto de Ciencia de Materiales de Madrid, CSIC, Cantoblanco
E28049 Madrid, Spain}

\author{R. Rold\'an}
\email[Electronic address: ]{rroldan@icmm.csic.es}
\affiliation{Instituto de Ciencia de Materiales de Madrid, CSIC, Cantoblanco
E28049 Madrid, Spain}

\author{E. Cappelluti}
\affiliation{Istituto dei Sistemi Complessi, CNR, U.O.S. Sapienza, v. dei
Taurini 19, 00185 Roma, Italy}

\author{F. Guinea}
\affiliation{Instituto de Ciencia de Materiales de Madrid, CSIC, Cantoblanco
E28049 Madrid, Spain}

\author{A. Fasolino}
\affiliation{Radboud University Nijmegen,Institute for Molecules and Materials,
NL-6525AJ Nijmegen, The Netherlands}

\author{M. I. Katsnelson}
\affiliation{Radboud University Nijmegen,Institute for Molecules and Materials,
NL-6525AJ Nijmegen, The Netherlands}

\date{\today}


\begin{abstract}
We investigate the thermodynamic properties and the lattice stability
of two-dimensional crystalline membranes, such as graphene and related
compounds, in the low temperature quantum regime $T\rra0$. A key
role is played by the anharmonic coupling between in-plane and out-of-plane lattice modes that, in the quantum limit, has very different consequences from those in the classical regime. 
The role of retardation, namely of the frequency dependence,  in the effective anharmonic interactions 
turns out to be crucial in the quantum regime.  We identify a crossover temperature, $T^{*}$,
between classical and quantum regimes, which is $\sim 70 - 90$ K for graphene.
Below $T^{*}$, the heat capacity and thermal expansion coefficient decrease as power laws with decreasing temperature, 
tending to zero for $T\rra0$ as required by the third law of thermodynamics.

\end{abstract}
\maketitle

\section{Introduction}

The study of the  mechanical and thermodynamic  properties of membranes is a problem of broad interest in physics, being at the interface of statistical mechanics, condensed matter and field theory. Recent experimental developments in growing and isolating single
layers of crystalline materials, such  as graphene, MoS$_{2}$,
WS$_{2}$, BN and similar materials \cite{NG05}, have revived the interest in the properties of crystalline membranes.  The very thermodynamic stability
of these two-dimensional (2D) crystals has been a matter of debate (for a review see
Refs.~\onlinecite{K12,KF12}). 
At the harmonic level,  the out-of-plane
modes and the in-plane modes are completely decoupled, and the contribution of both modes to 
the mean-square atomic displacement diverges. In this situation, anharmonic
coupling between them should be taken into account \cite{NPW04}. This coupling suppresses the out-of-plane instability 
but increases the in-plane instability. As a result, strictly speaking, no long-range-order  exists at finite temperature
in agreement with the Mermin-Wagner theorem. This fact implies that Bragg peaks are not infinitely sharp in the thermodynamic limit, as in 3D ideal crystals, but for stiff membranes such as graphene they are still narrow and centred at the regular reciprocal lattice sites. This situation defines precisely what we call 2D crystals~\cite{K12,KF12}.  
The anharmonic coupling between in-plane and out-of-plane modes is reflected in the onset of out-of-plane
ripples at any finite temperature \cite{FLK07}. The impact of corrugations on electronic transport, as well as on the mechanical properties of graphene, is a subject of intense investigation \cite{KG08,KC08,MO08,G09b,EK10,SGG11,MO10,CG10,GKM12,GP_2012,AG13}.  Ripples have been indeed measured
in atomically thick materials such as graphene \cite{MR07}
or MoS$_{2}$ \cite{BK11}, although it is still experimentally unclear
whether they are mostly thermal in nature or due to strain.

A crystalline membrane is a strongly anharmonic system, and these anharmonic effects must be taken into account beyond the usual quasi-harmonic approximation (which ignores phonon-phonon interaction). However,  the role of strong anharmonic effects in crystalline membranes has almost exclusively  been theoretically investigated in the  classical regime \cite{NP87,AL88,Paczuski_1988,Paczuski_1989,AL_1989,DR92,Bowick_1996,NPW04,KM09,LF09,BH10,G09c,ZRFK10,CD11,RFZK11,Hasselmann_2011,LK12,KN13,EKM14,KN_2014}. An exception to this are the recent works Refs.~\onlinecite{PacoDoussal_2013} and \onlinecite{Kats_2013}, where both quantum and anharmonic effects are taken into account. Nevertheless, Ref.~\onlinecite{Kats_2013} neglects some relevant anharmonic terms, while both Refs.~\onlinecite{PacoDoussal_2013} and  \onlinecite{Kats_2013} only partially take into account quantum fluctuations, neglecting effects of retardation in the interactions.

The limitation of the classical approaches to high temperatures makes them
unsuitable to investigate the stability and thermodynamic properties
of such compounds in the very low temperature regime, where quantum
effects are dominant. Even the temperature above which quantum fluctuations can be neglected and the classical analysis becomes valid has not been known until now. However, taking graphene as an example, a simple estimation of its Debye temperature for the out-of-plane mode gives us a value of the order of $1000$ K, which hints that quantum fluctuations should be relevant even at relatively high temperatures.  

A quantum analysis can be
easily performed at the harmonic level, but the
lack of coupling between in-plane and out-of-plane modes leaves
the system unstable. This is reflected, for instance, in the fact that, in the thermodynamic limit and at a quasi-harmonic level, a divergent
negative areal thermal expansion $\al_{V}=-\infty$ is obtained \cite{AGK12},
in violation of the third law of thermodynamics, which implies that
the thermal expansion should vanish at $T=0$. In Ref.~\onlinecite{AGK12},
the effects of the interaction between out-of-plane and in-plane modes
have been included simply by means of a reasonable infrared (IR) cutoff
in the available momentum space of the harmonic model. This approach does not solve the problem because it
leads to a temperature independent (up to logarithmic accuracy) thermal
expansion \cite{AGK12}. This difficulty   is due to the use of  the
classical anharmonic theory of a crystalline membrane, which cannot
be extrapolated to the $T\rra0$ limit.

Another important quantity is  the specific heat. The harmonic theory predicts that at very low temperature,
it should be dominated by the out-of-plane mode and, due to its quadratic
dispersion relation at long wavelength, the heat capacity should behave as $c_{p}\sim T$
\cite{ZPC08,Popov_2002,MM05,AGK12}. It is important to understand how this picture
changes once we include the effects of anharmonic  interactions, since we know that
they  drastically change the properties of
the out-of-plane modes \cite{NP87,AL88,DR92,KM09,ZKF09,ZRFK10}.

A robust description of the lattice properties
of 2D crystalline membranes, satisfying the third law
of thermodynamics in the low temperature limit and properly including
both quantum effects and the anharmonic coupling between in-plane
and out-of-plane modes, is still lacking. 

In this paper, we develop a theory for 
anharmonic crystalline membranes in the quantum regime. Toward this end,
we derive an effective quantum field theory governing the dynamics
of the out-of-plane modes. The in-plane modes can be integrated out rigorously leading to an effective retarded (i. e., frequency dependent) interaction between out-of-plane modes. Then, we study the effect of the anharmonicities
by computing the self-energy to first order
in perturbation theory, obtaining the corresponding corrections to the elastic constants. 
It is known that perturbation theory is insufficient
to describe the physics of the classical version of this
problem \cite{NPW04}, making it necessary to use some other techniques
such as $\epsilon=4-D$ expansion \cite{AL88} (where $D$ is the membrane dimension), the self-consistent screening approximation (SCSA) \cite{DR92}, or the non-perturbative renormalization group method \cite{KM09}. 
Therefore, we have further worked on the first step beyond perturbation theory, by using a one-loop self-consistent theory, without including any renormalization of the in-plane Lam\'e constants, which can be viewed as the generalization of the Nelson and Peliti approximation \cite{NP87} to the quantum regime. The present
work can therefore be seen as the first stage of the full description of quantum crystalline membranes. 
However, the perturbative
calculation is already useful to study two problems: (i) assess the
effect of retardation in the effective interaction between out-of-plane
modes; (ii) investigate the momentum space associated with anharmonic
effects by applying a Ginzburg criterion. Comparing the perturbative
calculation performed in the quantum regime at $T=0$, with the result
from the classical theory, allows us to determine
a cross over  temperature $T^{*}$, below which quantum effects become dominant.
Finally, we study the effect of anharmonicities on the thermal expansion and specific heat
of quantum crystalline membranes, solving the  contradiction with the third law of thermodynamics.

\section{Model}

We start our analysis from the standard continuum theory for crystalline
elastic membranes and thin plates \cite{LL59,NP87,AL88,CL03,NPW04,K12}. The lattice deformations
of the membrane are expressed in terms of an in-plane 2D vector displacement
field $\vec{u}$ and an out-of-plane (flexural) displacement field $h$.
We will use the imaginary time functional path integral formalism,
which is particularly convenient to study thermodynamic quantities.
The Euclidean action can be written as $\mathcal{S}\left[\vec{u},h\right]=\int_{0}^{\beta}d\tau\int d^{2}x\mathcal{L}\left[\vec{u},h\right]$,
 where $0<\tau<\beta$ is the imaginary time and $\beta=1/(k_{B}T)$
is the inverse temperature, 
with Lagrangian density 
\begin{equation}
\mathcal{L}\left[\vec{u},h\right]=\mathcal{L}_{h}^{0}\left[h\right]+\mathcal{L}_{u}^{0}\left[\vec{u}\right]+\mathcal{L}_{\text{int}}^{(3)}\left[\vec{u},h\right]+\mathcal{L}_{\text{int}}^{(4)}\left[h\right].\label{eq:EuclideanAction}
\end{equation}
Here, $\mathcal{L}_{h}^{0}\left[h\right]$ and $\mathcal{L}_{u}^{0}\left[\vec{u}\right]$
are the quadratic Lagrangian densities for the out-of-plane and in-plane
displacement fields, 
\begin{align}
\mathcal{L}_{h}^{0}\left[h\right] & =\frac{1}{2}\rho\dot{h}^{2}+\frac{1}{2}\ka\left(\pd^{2}h\right)^{2},\label{eq:harmonic_out}\\
\mathcal{L}_{u}^{0}\left[\vec{u}\right] & =\frac{1}{2}\rho\dot{\vec{u}}^{2}+\frac{1}{2}c^{ijkl}\pd_{i}u_{j}\pd_{k}u_{l},\label{eq:harmonic_in}
\end{align}
and $\mathcal{L}_{\text{int}}^{(3)}\left[\vec{u},h\right]$ and $\mathcal{L}_{\text{int}}^{(4)}\left[h\right]$
are anharmonic terms. $\mathcal{L}_{\text{int}}^{(3)}\left[\vec{u},h\right]$
contains cubic interactions between in-plane and out-of-plane modes,
and $\mathcal{L}_{\text{int}}^{(4)}\left[h\right]$ accounts for a
quartic local interaction for the out-of-plane field. Explicitly we
have 
\begin{align}
\mathcal{L}_{\text{int}}^{(3)}\left[\vec{u},h\right] & =\frac{1}{2}c^{ijkl}\pd_{i}u_{j}\left(\pd_{k}h\pd_{l}h\right),\label{eq:cubic_u-h}\\
\mathcal{L}_{\text{int}}^{(4)}\left[h\right] & =\frac{1}{8}c^{ijkl}\left(\pd_{i}h\pd_{j}h\right)\left(\pd_{k}h\pd_{l}h\right).\label{eq:quartic_on-site}
\end{align}
In the above expressions, $\rho$ is the mass density, $\ka$ is the bending rigidity, $c^{ijkl}=\lambda\de^{ij}\de^{kl}+\mu\left(\de^{ik}\de^{jl}+\de^{il}\de^{jk}\right)$
is the elastic moduli tensor, $\mu$ and $\la$ are Lam\'{e} coefficients \footnote{%
We use graphene as an example of a crystalline membrane. Typical parameters for single-layer graphene at $T=0$
are taken (see Refs.~\protect\onlinecite{FLK07,ZKF09}): $\mu=9.44$ eV \AA$^{-2}$, $\lambda=3.25$ eV \AA$^{-2}$ and
$\kappa=0.82$ eV. At $T=300$K we used the values: $\mu=9.95$ eV \AA$^{-2}$, $\lambda=2.57$ eV \AA$^{-2}$ and
$\kappa=1.1$ eV. Graphene has density $\rho/\hbar^2=1104$~eV$^{-1}$\AA$^{-4}$ and its lattice constant is given by $a=2.46$ \AA , from which we obtain a Debye momentum $q_D=\sqrt{8\pi/(3^{1/2}a^2)}=1.55$\AA$^{-1}$. 
}.
The latin indices ($i,j,...$) run over the spatial coordinates $x,\, y$,
and we use the convention where repeated indices are to be summed
over. In addition, we write $\dot{O}\equiv\pd O/\pd\tau$ ($O=h,\vec{u}$).

It is known that in the classical problem both anharmonic terms $\mathcal{L}_{\text{int}}^{(3)}\left[\vec{u},h\right]$ and $\mathcal{L}_{\text{int}}^{(4)}\left[h\right]$ are equally relevant \cite{NP87,AL88,AL_1989}. Inclusion of $\mathcal{L}_{\text{int}}^{(4)}\left[h\right]$ is also needed in order to make the Euclidean action bounded from below, and therefore to have a well defined ground state. Therefore, we keep both terms in the quantum theory. It is worthwhile noting that the term $\mathcal{L}_{\text{int}}^{(4)}\left[h\right]$ was not considered in Ref.~\onlinecite{Kats_2013}. 

The partition function is written as the
functional integral $Z=\int D\left[\vec{u},h\right]\exp\left(-\mathcal{S}\left[\vec{u},h\right]\right)$.
The classical treatment formally corresponds to neglecting all kinetic terms in
the Lagrangian density. To be able to reach the low temperature limit, it is
necessary to take into account quantum fluctuations of the fields
$h$ and $\vec{u}$, by retaining  the kinetic terms
in the Euclidean action. It is convenient to express the fields in
Fourier components 
\begin{equation}
O(\vec{x},\tau)=\frac{1}{\sqrt{\beta V}}\sum_{\boldsymbol{k}}O_{\boldsymbol{k}}e^{i\vec{k}\cdot\vec{x}}e^{-ik_{n}\tau},
\end{equation}
where $k_{n}=2\pi n/\beta$, with $n\in\mathbb{Z}$, are bosonic Matsubara
frequencies, $V$ is the area of the undistorted membrane and we have
used the shorthand notation $\bm{k}=\left(ik_{n},\vec{k}\right)$
with $\sum_{\bm{k}}=\sum_{ik_{n},\vec{k}}$. We will later see that the thermodynamic quantities we are interested in can be expressed via the two-point correlation functions (propagators) $G_{\bm{k}}=\left\langle h_{\boldsymbol{k}}h_{-\boldsymbol{k}}\right\rangle $
and $D_{\bm{q}}^{ij}=\left\langle u_{\boldsymbol{q}}^{i}u_{-\boldsymbol{q}}^{j}\right\rangle $,
where $\left\langle O\right\rangle =Z^{-1}\int D\left[\vec{u},h\right]O\exp\left(-\mathcal{S}\left[\vec{u},h\right]\right)$.
At the level of the harmonic theory, the correlation functions are
given by 
\begin{align}
G_{\bm{k}}^{0} & =\left(-\rho\left(ik_{n}\right)^{2}+\rho\omega_{k,F}^{2}\right)^{-1},\label{eq:bare_out}\\
D_{\bm{q}}^{0,L} & =\left(-\rho\left(iq_{n}\right)^{2}+\rho\omega_{q,L}^{2}\right)^{-1},\label{eq:bare_in_L}\\
D_{\bm{q}}^{0,T} & =\left(-\rho\left(iq_{n}\right)^{2}+\rho\omega_{q,T}^{2}\right)^{-1},\label{eq:bare_in_T}
\end{align}
where we have split $D_{\bm{q}}^{ij,0}$ in its longitudinal ($L$) and transverse ($T$)
components with respect to the vector $\vec{q}$. The bare dispersion relations for the
flexural ($F$) and  in-plane longitudinal/transverse modes are,
respectively, $\omega_{k,F}=\sqrt{\kappa/\rho}k^{2}$ and  $\omega_{q,L/T}=c_{L/T}q$, with $c_{L}=\sqrt{(\lambda+2\mu)/\rho}$ and $c_{T}=\sqrt{\mu/\rho}$. When anharmonic effects are
taken into account, the two point correlation functions are given by the Dyson
equations, 
\begin{align}
G_{\bm{k}}^{-1} & =\left(G_{\bm{k}}^{0}\right)^{-1}+\Sigma_{\bm{k}},\label{eq:full_out}\\
\left(D_{\bm{q}}^{L}\right)^{-1} & =\left(D_{\bm{q}}^{0,L}\right)^{-1}+\mathcal{P}_{\bm{q}}^{L},\\
\left(D_{\bm{q}}^{T}\right)^{-1} & =\left(D_{\bm{q}}^{0,T}\right)^{-1}+\mathcal{P}_{\bm{q}}^{T},
\end{align}
where $\Sigma_{\bm{k}}$ is the self-energy for the out-of-plane mode
and $\mathcal{P}_{\bm{q}}^{L/T}$ is the self-energy for the in-plane
longitudinal/transverse mode. Since the Euclidean action \eqref{eq:EuclideanAction}
is quadratic in the field $\vec{u}$, the latter can be integrated out exactly,
so that we are left with an effective theory only involving the flexural field
$h$. Doing this (details given in Appendix~\ref{sec:Appendix_eff_action}),
the effective theory is described by the action 
\begin{multline}
\mathcal{S}_{\text{eff}}\left[h\right]=\frac{1}{2}\sum_{\boldsymbol{k}}\left(-\rho\left(ik_{n}\right)^{2}+\kappa k^{4}\right)h_{\boldsymbol{k}}h_{-\boldsymbol{k}}\\
+\frac{1}{8\beta V}\sum_{\boldsymbol{k},\boldsymbol{p},\boldsymbol{q}\neq0}R_{\bm{q}}^{ijkl}\left(k+q\right)_{i}k_{j}\left(p-q\right)_{k}p_{l}\\
\times h_{\boldsymbol{k}+\boldsymbol{q}}h_{-\boldsymbol{k}}h_{\boldsymbol{p}-\boldsymbol{q}}h_{-\boldsymbol{p}}\label{eq:S_eff}
\end{multline}
where $R_{\bm{q}}^{ijkl}$ is the effective interaction tensor between
out-of-plane modes that takes into account both interaction channels:
the quartic local interaction \eqref{eq:quartic_on-site} and the
in-plane mode mediated interaction due to the cubic interaction \eqref{eq:cubic_u-h}.
Just like in the classical theory, the $\bm{q}=0$ component was excluded
from the interaction term (see Appendix~\ref{sec:Appendix_eff_action}
and Ref.~\onlinecite{NPW04}). The tensor $R_{\bm{q}}^{ijkl}$ obeys the
same symmetries of the elastic moduli  tensor, namely $R_{\bm{q}}^{ijkl}=R_{\bm{q}}^{jikl}=R_{\bm{q}}^{klij}$.
For a physical 2D membrane, $R_{\bm{q}}^{ijkl}$ has four independent
components, which are most conveniently written in the basis defined
by the momentum vector $\vec{q}$, $\left\{ \hat{e}_{\parallel},\hat{e}_{\perp}\right\} $,
with $\hat{e}_{\parallel}=\left(q_{x},q_{y}\right)/\left|\vec{q}\right|$
and $\hat{e}_{\perp}=\left(-q_{y},q_{x}\right)/\left|\vec{q}\right|$.
These are given by \footnote{%
These terms are also given, in a somewhat different form, in Ref.~\protect\onlinecite{PacoDoussal_2013}.
} 
\begin{align}
R_{\bm{q}}^{\perp\perp\perp\perp,\text{cl}} & =\frac{4\mu\left(\la+\mu\right)}{\left(\la+2\mu\right)}\label{eq:R_tttt}\\
R_{\bm{q}}^{\perp\perp\perp\perp,\text{qt}} & =-\rho\left(iq_{n}\right)^{2}\frac{\la^{2}}{\la+2\mu}D_{\bm{q}}^{0,L}\label{eq:R_tttt,qt}\\
R_{\bm{q}}^{\parallel\parallel\parallel\parallel} & =-\rho\left(iq_{n}\right)^{2}(\la+2\mu)D_{\bm{q}}^{0,L},\label{eq:R_pppp}\\
R_{\bm{q}}^{\perp\perp\parallel\parallel} & =-\rho\left(iq_{n}\right)^{2}\la D_{\bm{q}}^{0,L},\label{eq:R_ttpp}\\
R_{\bm{q}}^{\perp\parallel\perp\parallel} & =-\rho\left(iq_{n}\right)^{2}\mu D_{\bm{q}}^{0,T},\label{eq:R_tptp}
\end{align}
where we have written the $R_{\bm{q}}^{\perp\perp\perp\perp}$ component
as $R_{\bm{q}}^{\perp\perp\perp\perp}=R_{\bm{q}}^{\perp\perp\perp\perp,\text{cl}}+R_{\bm{q}}^{\perp\perp\perp\perp,\text{qt}}$.
The term $R_{\bm{q}}^{\perp\perp\perp\perp,\text{cl}}$ is frequency independent, and coincides with the 2D Young modulus. This term 
is already present in the classical statistical mechanics problem
\cite{NP87,NPW04}, and this is why we denote it by the superscript cl, from {\it classical}. The remaining interaction terms, $R_{\bm{q}}^{\perp\perp\perp\perp,\text{qt}}$,
$R_{\bm{q}}^{\parallel\parallel\parallel\parallel}$, $R_{\bm{q}}^{\perp\perp\parallel\parallel}$
and $R_{\bm{q}}^{\perp\parallel\perp\parallel}$, are new terms that
do not occur in the classical theory, for which reason we will refer
to them as {\it quantum}. The {\it quantum} terms \eqref{eq:R_tttt,qt}-\eqref{eq:R_tptp}
all have the same structure, depending on the Matsubara frequency
and become zero for $iq_{n}=0$. Therefore the term $R_{\bm{q}}^{ijkl}\left(k+q\right)_{i}k_{j}\left(p-q\right)_{k}p_{l}$
that appears in \eqref{eq:S_eff} can be written as
\begin{equation}
R_{\bm{q}}^{ijkl}\left(k+q\right)_{i}k_{j}\left(p-q\right)_{k}p_{l}=\sum_{M}g_{\vec{k},\vec{q},\vec{p}}^{M}R_{\bm{q}}^{M},\label{eq:scalar_int}
\end{equation}
with $M$ running over $\left\{ ^{\perp\perp\perp\perp},^{\parallel\parallel\parallel\parallel},^{\perp\perp\parallel\parallel},^{\perp\parallel\perp\parallel}\right\} $, $g_{\vec{k},\vec{p},\vec{q}}^{M}$ given by
\begin{align}
g_{\vec{k},\vec{p},\vec{q}}^{\perp\perp\perp\perp} & =k^{2}p^{2}\sin^{2}\theta_{k,q}\sin^{2}\theta_{p,q},\label{eq:g_tttt}\\
g_{\vec{k},\vec{p},\vec{q}}^{\parallel\parallel\parallel\parallel} & =kp\cos\theta_{k,q}\cos\theta_{p,q}\times\nonumber \\
 & \times\left(k\cos\theta_{k,q}+q\right)\left(p\cos\theta_{p,q}-q\right),\\
g_{\vec{k},\vec{p},\vec{q}}^{\perp\perp\parallel\parallel} & =k^{2}p\sin^{2}\theta_{k,q}\cos\theta_{p,q}\left(p\cos\theta_{p,q}-q\right)\nonumber \\
 & +p^{2}k\sin^{2}\theta_{p,q}\cos\theta_{k,q}\left(k\cos\theta_{p,q}+q\right),\\
g_{\vec{k},\vec{p},\vec{q}}^{\perp\parallel\perp\parallel} & =kp\sin\theta_{k,q}\sin\theta_{p,q}\times\nonumber \\
 & \times\left(2k\cos\theta_{k,q}+q\right)\left(2p\cos\theta_{p,q}-q\right),\label{eq:g_tptp}
\end{align}
where $\theta_{k,q}$ and $\theta_{p,q}$ are, respectively, the angle
between $\vec{k}$ and $\vec{q}$, and the angle between $\vec{p}$
and $\vec{q}$. It is important to emphasize that both the quartic local
interaction \eqref{eq:quartic_on-site} and the in-plane mode mediated
interaction \eqref{eq:cubic_u-h} contribute to all the interaction
terms \eqref{eq:R_tttt}-\eqref{eq:R_tptp}. In Appendix~\ref{sec:Appendix_eff_action}
we show the individual contributions from the quartic local interaction
and from the in-plane mediated channel.

\section{Classical versus Quantum regimes}

\begin{figure}
\centering{}
\includegraphics[width=8cm]{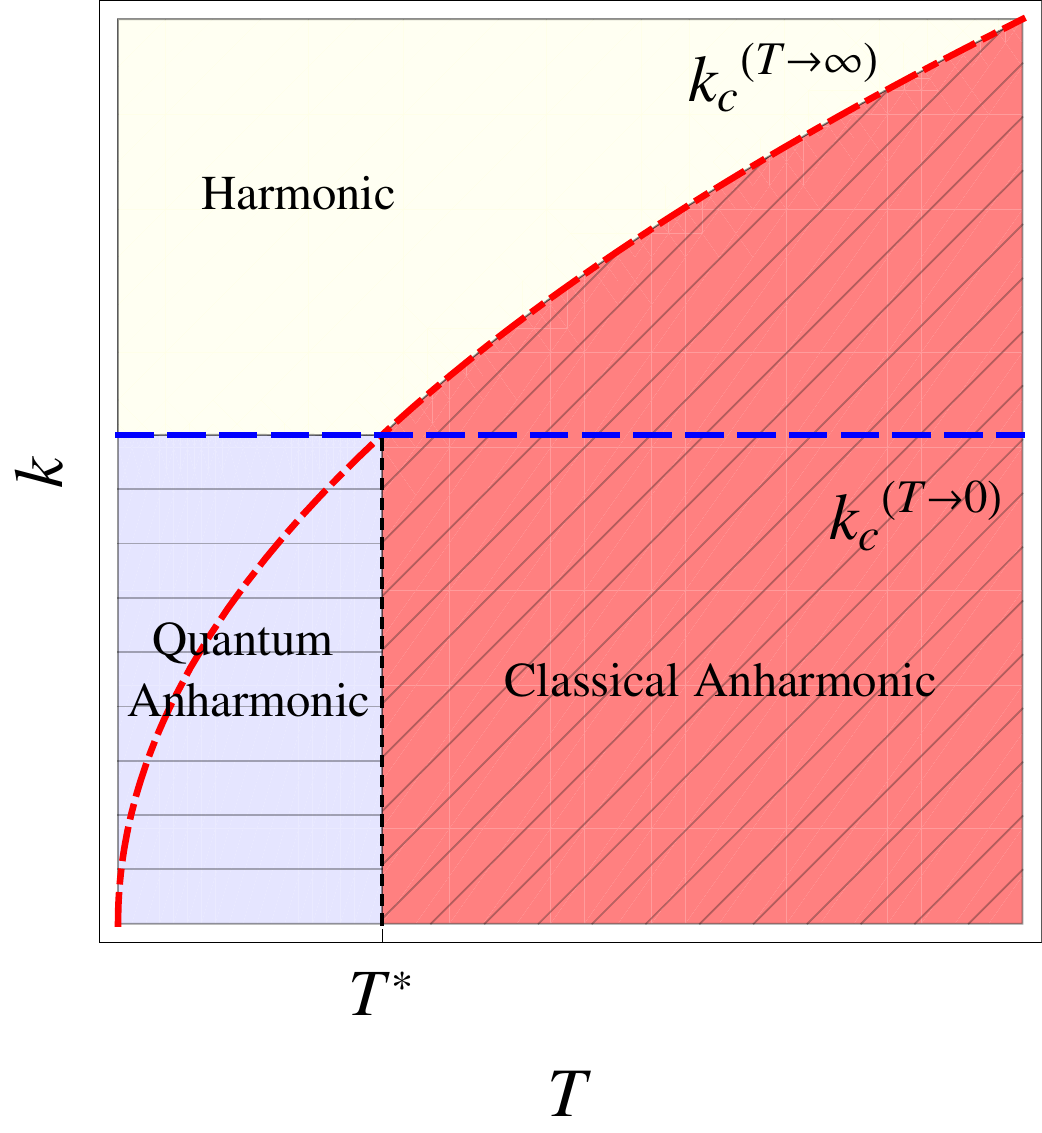} 
\caption{\label{fig:Phase-diagram}Phase diagram in $\left(T,k\right)$ space
for the harmonic/anharmonic and quantum/classical regimes. The harmonic-to-anharmonic
crossover momenta $k_{c}^{(T\rra\infty)}$ \eqref{eq:kc_high} and
$k_{c}^{(T\rra0)}$\eqref{eq:kc_low} are displayed by the dot-dashed red and dashed blue lines, respectively. The vertical
line shows the classical-to-quantum crossover temperature $T^{*}$.
In the region labelled \textit{Harmonic}, anharmonic effects are weak and the membrane is
nearly harmonic. In the region labelled \textit{Quantum Anharmonic}, anharmonic effects dominate and the
main contribution comes from the $T\rra0$ {\it quantum} terms, Eq.~\eqref{eq:Sigma_T_0}.
In the region labelled \textit{Classical Anharmonic}, anharmonic effects dominate and the main contribution
comes from the $T\rra\infty$ {\it classical} term, Eq.~\eqref{eq:Sigma_t_inf}. }
\end{figure}

In the high temperature limit, $T\rra\infty$, the main contribution
to the self-energy of the out-of-plane mode comes from the only interaction
term that occurs in the classical theory, $R_{\bm{q}}^{\perp\perp\perp\perp,\text{cl}}$
(see Appendix~\ref{sec:Appendix_self_energy_flexural} for an analysis
of the remaining terms). To first order in perturbation theory, after
an analytic continuation to real frequencies $ik_{n}\rra\omega+i0^{+}$,
we obtain the well known result \cite{NPW04} 
\begin{equation}
\lim_{T\rra\infty}\re\Sigma_{k}(\omega)\simeq\frac{4\mu\left(\la+\mu\right)}{\left(\la+2\mu\right)\kappa}\frac{3k_{B}T}{16\pi}k^{2},\label{eq:Sigma_t_inf}
\end{equation}
which is frequency independent. In the low temperature limit,
$T\rra0$, to first order in perturbation theory, the main contribution
to the on-shell self-energy ($\omega=\omega_{k,F}$) in the physically relevant
long wavelength limit does not result 
from the $R_{\bm{q}}^{\perp\perp\perp\perp}$ component,
but from the quantum terms $R_{\bm{q}}^{\parallel\parallel\parallel\parallel}$
and $R_{\bm{q}}^{\perp\parallel\perp\parallel}$, due to the factors
$g_{\vec{k},-\vec{k},\vec{q}}^{\parallel\parallel\parallel\parallel}$ and $g_{\vec{k},-\vec{k},\vec{q}}^{\parallel\parallel\perp\parallel}$
that behave as $\sim k^{2}$ for $k\rra0$. In this limit, we obtain
(see Appendix~\ref{sec:Appendix_self_energy_flexural}) 
\begin{align}
\lim_{T\rra0}\re\Sigma_{k}\left(\omega_{k,F}\right) & \simeq\frac{\hbar\kappa^{1/2}k^{2}}{8\pi\rho^{1/2}}\times\nonumber \\
 & \times\left[q_{L}^{4}f\left(\frac{\Lambda}{q_{L}}\right)+q_{T}^{4}f\left(\frac{\Lambda}{q_{T}}\right)\right],\label{eq:Sigma_T_0}
\end{align}
where we have imposed an ultraviolet (UV) momentum cutoff $\Lambda$ (which should be of the order of the Debye momentum, $q_D$), and we have defined
the function 
\begin{equation}\label{Eq:f}
f\left(x\right)=\frac{1}{2}x\left(x-2\right)+\log\left(1+x\right),
\end{equation}
and introduced the two momentum scales \footnote{
For graphene values at $T=0$ we have
$q_L \simeq 5.2\text{\AA}^{-1}$ and $q_{T}\simeq 3.4\text{\AA}^{-1}$.
Therefore, the momentum scales $q_{L/T}$ are actually larger than
graphene's Debye momentum.
}
\begin{eqnarray}\label{Eq:qLT}
q_{L}&=&\sqrt{\left(\la+2\mu\right)/\kappa}\nonumber\\
q_{T}&=&\sqrt{\mu/\kappa}.
\end{eqnarray} %
Notice [Eqs. \eqref{eq:Sigma_T_0} and \eqref{Eq:f}] that besides logarithmic UV divergences, we have also obtained
power law divergences. Careful inspection allows us to see that the
main $\Lambda^{2}$ divergence comes from the quartic local interaction,
$\mathcal{L}_{\text{int}}^{(4)}\left[h\right]$, while the $\Lambda$ and $\log(\Lambda)$ divergences come from the interaction of in-plane with out-of-plane modes, $\mathcal{L}_{\text{int}}^{(3)}\left[\vec{u}, h\right]$. 
To first order in perturbation theory,
the dispersion relation, $\Om_{k}$, of the physical
excitations is given by $\rho\Omega_{k}^{2}=\kappa k^{4}+\re\Sigma_{k}\left(\omega_{k,F}\right)$.
Note that, for $k\rightarrow 0$, $\Omega_{k} \sim k$ instead of $\omega_{k,F} \sim k^2$.
It can be checked that, in the long wavelength
limit, the result from Eq.~\eqref{eq:Sigma_T_0} is
the same as that obtained by setting $\omega=0$ in $\Sigma_k(\omega)$.  This tells us that, for physical excitations, the frequency dependence
of the self-energy can be neglected in the low temperature and long wavelength
limit. We can estimate, in both the high temperature and low temperature limits, the momentum scale, $k_{c}$, below which anharmonic
effects become dominant, as determined by the Ginzburg criterion~\cite{NPW04} $\Sigma_{k_{c}}\left(\omega_{k_{c},F}\right)=\kappa k_{c}^{4}$.
By doing such analysis, we obtain \footnote{There is a factor of $1/\sqrt{2}$ missing in the expression for $ k_c^{ (T\rra\infty )}$ presented in Refs.~\protect\onlinecite{FLK07,ZRFK10,RFZK11}
}

\begin{align}
k_{c}^{(T\rra\infty)} & \simeq \sqrt{\frac{3k_{B}T}{16\pi}\frac{4\mu\left(\la+\mu\right)}{\kappa^{2}\left(\la+2\mu\right)}},\label{eq:kc_high}\\
k_{c}^{(T\rra0)} & \simeq \sqrt{\frac{\hbar}{8\pi\rho^{1/2}\kappa^{1/2}}}\left[q_{L}^{4}f\left(\frac{\Lambda}{q_{L}}\right)+q_{T}^{4}f\left(\frac{\Lambda}{q_{T}}\right)\right]^{1/2}.\label{eq:kc_low}
\end{align}
For typical graphene values we obtain $k_{c}^{(T\rra\infty)} \simeq 0.17\text{\AA}^{-1}$ at $T=300$K (in agreement with what was found in Ref.~\onlinecite{LF09}), and setting $\Lambda=q_{D}$,
we obtain $k_{c}^{(T\rra0)}\simeq 0.1\,\text{\AA}^{-1}$ at $T=0$.  It is useful
to write approximate expressions for $k_{c}$ in the limit $T\rra0$
when $\Lambda/q_{T/L}\gg1$ and $\Lambda/q_{T/L}\ll1$. Expanding
the function $f(x)$, we obtain the following approximate expressions:
\begin{equation}
k_{c}^{(T\rra0)}\simeq\begin{cases}
\left(\frac{\hbar}{16\pi\sqrt{\rho\kappa}}\frac{\la+3\mu}{\kappa}\right)^{1/2}\Lambda, &\Lambda\gg q_{T/L},\\
\left(\frac{\hbar}{24\pi\sqrt{\rho\kappa}}\right)^{1/2}\left(q_{L}+q_{T}\right)^{1/2}\Lambda^{3/2}. &\Lambda\ll q_{T/L}.
\end{cases}
\end{equation}
To determine the actual importance of the anharmonic effects in suspended samples of crystalline membranes, one has to compare the anharmonic scale $k_{c}$, with the minimum momentum allowed by the finite size of the sample $\sim 1 /L$ and with the momentum scale due to residual strains $\sim q_{L} \bar{u}$, where $\bar{u}$ is the strain \cite{CG10}.
We can also estimate the temperature, $T^{*}$, at which  the
crossover from the classical to the quantum regime occurs, by equating
$\lim_{T\rra\infty}\re\Sigma_{k}(\omega_{k,F})=\lim_{T\rra0}\re\Sigma_{k}(\omega_{k,F})$.
Since, in both regimes, the leading contribution to the self-energy
goes like $k^{2}$, this is equivalent to comparing $k_{c}^{(T\rra\infty)}$
with $k_{c}^{(T\rra0)}$. We obtain 
\begin{align}
T^{*} & \simeq \frac{2\hbar}{3k_{B}\rho^{1/2}\kappa^{1/2}}\frac{\left(\la+2\mu\right)\kappa^{2}}{4\mu\left(\la+\mu\right)}\times\nonumber \\
 & \times\left[q_{L}^{4}f\left(\frac{\Lambda}{q_{L}}\right)+q_{T}^{4}f\left(\frac{\Lambda}{q_{T}}\right)\right],
\end{align}
a quantity that depends on the UV cutoff $\Lambda$. Expanding once
again the function $f(x)$ we obtain 
\begin{equation}
T^{*} \simeq \begin{cases}
\frac{\hbar\kappa^{1/2}}{3k_{B}\rho^{1/2}}\frac{\left(\la+2\mu\right)\left(\la+3\mu\right)}{4\mu\left(\la+\mu\right)}\Lambda^{2}, &\Lambda\gg q_{T/L},\\
\frac{2\hbar\kappa}{9k_{B}\rho^{1/2}}\frac{\la+2\mu}{4\mu\left(\la+\mu\right)}\left(\sqrt{\la+2\mu}+\sqrt{\mu}\right)\Lambda^{3}, &\Lambda\ll q_{T/L}.
\end{cases}
\end{equation}
For typical graphene values, setting $\Lambda=q_{D}$, we obtain a value of
$T^{*} \sim 70 - 90 \text{ K}$ (depending on the values we take for the elastic constants, which depend themselves on the temperature \cite{FLK07,ZKF09}). Below this temperature, the contribution
to the self-energy from the quantum interaction terms $R_{\bm{q}}^{\parallel\parallel\parallel\parallel}$
and $R_{\bm{q}}^{\perp\parallel\perp\parallel}$ should become dominant.
Fig.~\ref{fig:Phase-diagram} shows the different regions in the $\left(T,k\right)$
space where anharmonic and quantum effects give the main contribution.

It is interesting to notice that both in the classical and in the
quantum regime, the self-energy has the same $k^{2}$ dependence,
with negligible frequency dependence. However, it must be kept in
mind that the main contributions to the self-energy have very different
origins. In fact, it can be checked (see Appendix~\ref{sec:Appendix_self_energy_flexural})
that the contribution of the interaction term $R_{\bm{q}}^{\perp\perp\perp\perp,\text{cl}}$
for $T\rra0$ goes as $\Sigma_{k}^{\text{cl}}\propto k^{4}\log\left(\Lambda/k\right)$.
In this $T\rra0$ limit, it is clear that the contribution to the
self-energy from $R_{\bm{q}}^{\parallel\parallel\parallel\parallel}$
and $R_{\bm{q}}^{\perp\parallel\perp\parallel}$ \eqref{eq:Sigma_T_0},
dominates over the contribution from $R_{\bm{q}}^{\perp\perp\perp\perp,\text{cl}}$
at small momenta. If we would extend the result \eqref{eq:Sigma_T_0}
to large $k$, going beyond its long wavelength validity region, 
it is clear that it would also dominate over the contribution from
$R_{\bm{q}}^{\perp\perp\perp\perp,\text{cl}}$, for large enough $k$.
One could think that there might exist an intermediate momentum region
where $\Sigma_{k}^{\text{cl}}$ actually dominates over the term \eqref{eq:Sigma_T_0}.
It can be checked however  that for usual graphene values Eq.~\eqref{eq:Sigma_T_0}
always dominates and that increasing the bending rigidity, $\kappa$, (i.e. to account for stiffer crystalline membranes as single-layer MoS$_2$)
makes this dominance even stronger. 

The next step to go beyond first order perturbation theory for $T\rra0$,
is to perform a simple self-consistent calculation in the spirit of
what was done by Nelson and Peliti in Ref.~\onlinecite{NP87}. First, we notice that in first
order perturbation theory, the in-plane mode propagator has only
logarithmic corrections due to anharmonic effects. This is
a much weaker effect than for the out-of-plane phonons
and therefore we will ignore it. Furthermore, we also neglect the frequency
dependence of the out-of-plane self-energy and write the full
out-of-plane correlation function as 
$G_{\bm{k}}\simeq\left(-\rho\left(ik_{n}\right)^{2}+\rho\omega_{k,F}^{2}+\kappa k_{c}^{\eta}k^{4-\eta}\right)^{-1}$,
where we have written the self-energy as $\Sigma_{k}=\kappa k_{c}^{\eta}k^{4-\eta}$,
with $k_{c}$ the harmonic-to-anharmonic crossover momentum, and $\eta$ a characteristic exponent, both  
to be determined in a self-consistent way. The main contribution to
the self-energy in the long wavelength limit still comes from $R_{\bm{q}}^{\parallel\parallel\parallel\parallel}$
and $R_{\bm{q}}^{\perp\parallel\perp\parallel}$, and since the
factors $g_{\vec{k},-\vec{k},\vec{q}}^{\parallel\parallel\parallel\parallel}$ and $g_{\vec{k},-\vec{k},\vec{q}}^{\perp\parallel\perp\parallel}$ behave as $\sim k^{2}$ for $k\rra0$, we obtain a value of $\eta=2$, in agreement with first order perturbation theory. This is an important result which justifies the use of first order perturbation theory contrary to  the classical regime, where the {\it perturbative} $\eta=2$ exponent is changed to $\eta=1$ when the same kind of self-consistent calculation is performed \cite{NP87}. The present result of $\eta=2$ indicates that quantum anharmonic effects act as an effective positive external strain, which contributes to the stabilization of the 2D phase of the membrane (see also Ref. \onlinecite{RFZK11}). 

Furthermore, the corresponding $k_{c}$ in this self-consistent approximation is determined by the condition
\begin{align}
k_{c}^{2} & =\frac{\left(\la+2\mu\right)\hbar k^{2}}{8\pi\rho^{1/2}\kappa^{3/2}}\int\frac{dqq^{3}}{\sqrt{q^{4}+k_{c}^{2}q^{2}}+q_{L}q}\nonumber \\
 & +\frac{\mu\hbar k^{2}}{8\pi\rho^{1/2}\kappa^{3/2}}\int\frac{dqq^{3}}{\sqrt{q^{4}+k_{c}^{2}q^{2}}+q_{T}q}.\label{eq:kc_SSC}
\end{align}
Notice that the integral over $q$ is convergent 
in the $q\rra0$ limit and does not depend on $k$. In the classical
theory, instead, the integral is singular in $k$ as $k^{\eta-2}$ and
$0<\eta\leq2$ \cite{NP87}. In fact, in the integral \eqref{eq:kc_SSC},
the term $q_{L}q$ dominates the denominator of the integrand for
small $q$, while $q^{2}$ dominates for large $q$. Therefore, the
term $k_{c}^{2}q^{2}$ will only contribute for intermediate values
of $q$ and the integral should be weakly dependent on $k_{c}$. Performing
the integral over momentum we obtain
\begin{equation}
k_{c}^{2}=\frac{\hbar}{8\pi\rho^{1/2}\kappa^{1/2}}\left[q_{L}^{4}F\left(\frac{\Lambda}{q_{L}},\frac{k_{c}}{q_{L}}\right)+q_{T}^{4}F\left(\frac{\Lambda}{q_{T}},\frac{k_{c}}{q_{T}}\right)\right],\label{eq:SelfConsistent_kc}
\end{equation}
where we have defined the function 
\begin{align}
F\left(x,y\right) & =\frac{1}{2}x\left(\sqrt{x^{2}+y^{2}}-2\right)+\frac{1}{2}\left(2-y^{2}\right)\sinh^{-1}\left(\frac{x}{y}\right)\nonumber \\
 & +\sqrt{1-y^{2}}\tanh^{-1}\left(\frac{x}{\sqrt{1-y^{2}}}\right)\nonumber \\
 & -\sqrt{1-y^{2}}\tanh^{-1}\left(\frac{x}{\sqrt{\left(1-y^{2}\right)\left(x^{2}+y^{2}\right)}}\right).
\end{align}
The function $F(x,y)$ reduces to $f(x)$ in the limit of $y\rra0$.
Solving the self-consistent condition \eqref{eq:SelfConsistent_kc}
for $k_{c}$ we obtain a value that, for typical graphene parameters, is nearly
unchanged with respect to the perturbative result $k_{c}\approx 0.1\,\text{\AA}^{-1}$.
The relative difference between the perturbative and self-consistent
value is of the order of $10^{-4}$. 

We are now in a position to calculate thermodynamics quantities as
the thermal expansion, $\alpha_{V}$, and the specific heat, $c_{p}$, in the low temperature quantum regime, which will be the focus of the next section.

\section{Thermodynamic quantities}

\subsection{Thermal expansion}

The areal thermal expansion coefficient is defined as 
\begin{equation}
\alpha_{V}=\frac{1}{V}\left(\frac{\pd\Delta V}{\pd T}\right)_{p},
\end{equation}
where $\Delta V$ is the change in the area of the membrane (to be
understood as the area of the membrane projected onto the reference $x-y$ plane),
$V$ is the area of the undistorted membrane, and the index $p$ indicates that
the process occurs at constant pressure/stress. Recalling that the relative
change of area is given by $\Delta V/V=\left\langle \pd_{i}u_{i}\right\rangle $,
the thermal expansion can be most efficiently computed by adding to the Euclidean Lagrangian \eqref{eq:EuclideanAction}
an extra term of the form $\mathcal{L}_{\text{\ensuremath{\sigma}}}\left[\vec{u}\right]=-\sigma\pd_{i}u_{i}$,
that describes an externally applied homogeneous and isotropic stress $\sigma$
to the membrane. The relative expansion at zero external stress can
therefore be obtained from
\begin{equation}
\frac{\Delta V}{V}=\frac{1}{\beta V}\frac{\pd}{\pd\sigma}\log Z\left[\sigma\right]\biggr|_{\sigma=0},
\end{equation}
where $Z\left[\sigma\right]=\int D\left[\vec{u},h\right]\exp\left(-\mathcal{S}\left[\vec{u},h\right]-\mathcal{S}_{\sigma}\left[\vec{u}\right]\right)$,
with $\mathcal{S}_{\sigma}\left[\vec{u}\right]=-\sigma\int_{0}^{\beta}d\tau\int d^{2}x\pd_{i}u_{i}$.
Notice that $\sigma$ couples to the $\bm{q}=0$ component of $\pd_{i}u_{i}$.
Therefore, we can perform a shift of $\pd_{i}u_{j}$ in the functional
path integral $\pd_{i}u_{j}\rra\pd_{i}u_{j}+c_{ijkl}^{-1}\de^{kl}\sigma$,
where $c_{ijkl}^{-1}$ is the inverse of the
elastic moduli tensor), and cancel the linear term in $\pd_{i}u_{i}$ of
$\mathcal{S}_{\sigma}\left[\vec{u}\right]$ at the expense of generating
two new terms in the Euclidean Lagrangian density: (i) one of the
form$-\sigma^{2}c_{iijj}^{-1}/2$, which does not lead to any thermal
expansion, (ii) and another term of the form $\sigma\pd_{i}h\pd_{i}h/2$,
which is the term responsible for the thermal expansion. Therefore
we can write the thermal expansion at zero external stress as \cite{AGK12}
\begin{equation}
\alpha_{V}=-\frac{1}{2}\frac{\pd}{\pd T}\left(\frac{1}{\beta V}\sum_{\bm{k}}k^{2}G_{\bm{k}}\right)\label{eq:thermal_exp}.
\end{equation}
Replacing in Eq.~\eqref{eq:thermal_exp} the full out-of-plane correlation function \eqref{eq:full_out}  by the bare, harmonic, one \eqref{eq:bare_out} is equivalent to performing a quasi-harmonic treatment of the thermal expansion of a membrane \cite{AGK12}. In the quasi-harmonic approximation one obtains
\begin{equation}
\alpha_{V}^{\text{qh}}=-\frac{\hbar^{2}}{8\rho k_{B}T^{2}}\int_{q_{\text{min}}}^{\Lambda}\frac{d^{2}k}{\left(2\pi\right)^{2}}\frac{k^{2}}{\sinh^{2}\left(\hbar\omega_{k,F}/(2k_{B}T)\right)},\label{eq:expansion_qh}
\end{equation}
where the UV momentum cutoff is of the order of the inverse lattice spacing, $\Lambda \sim 1/a$, and the IR cutoff is of the order of the inverse of the membrane size, $q_{\text{min}} \sim 1/L$. For not too high temperatures, we can safely take the limit $\Lambda\rra\infty$ in Eq.~\eqref{eq:expansion_qh}, but the integral is divergent for $q_{\text{min}}\rra0$. Keeping $q_{\text{min}}$ finite, we can perform the integration analytically. In the quasi-harmonic approximation, the thermal expansion is given by $\alpha_{V}^{\text{qh}}=-k_{B}\mathcal{I}(t)/\left(8\pi\kappa\right)$, where
\begin{equation}
\mathcal{I}(t)=\int_{\frac{1}{2t}}^{\infty}\frac{dxx}{\sinh(x)^{2}}=\frac{1}{2t}\coth\left(\frac{1}{2t}\right)-\log\left[2\sinh\left(\frac{1}{2t}\right)\right],\label{eq:qh_exp_int}
\end{equation}
and $t=\left(k_{B}T\rho^{1/2}\right)/\left( \hbar \kappa^{1/2}q_{\text{min}}^2\right)$. The fact that ignoring anharmonic effects makes the membrane unstable, is reflected in that the limits $T\rra0$ and $L\rra\infty$ do not commute. As a matter of fact for $t\gg 1$ we have $\mathcal{I}(t)\sim 1+\log\left(t \right) $ while for $t\ll 1$ we have $\mathcal{I}(t)\sim e^{-1/t}/t$. Therefore, in the quasi-harmonic approximation the thermal expansion depends crucially on the size of the membrane even for $L\rra\infty$. This fact is important when interpreting numerical results for the thermal expansion of membrane like materials based on the quasi-harmonic theory. In these numerical calculations the thermal expansion is calculated by (finite difference) differentiation of the temperature dependence
of the lattice parameter calculated for finite size supercells \cite{MM05,Kthesis_2010,KA_2011}. We will now see how going beyond the quasi-harmonic approximation will make the limits $T\rra0$ and $L\rra\infty$ commute so that the thermodynamic limit can be taken without ambiguity and at the same time satisfying the third law of thermodynamics. Taking into account the results from Eqs. \eqref{eq:Sigma_t_inf} and
\eqref{eq:Sigma_T_0}, we neglect the frequency dependence of the 
self-energy, writing the full out-of-plane correlation function as $G_{\boldsymbol{k}}^{-1}=-\rho\left(ik_{n}\right)^{2}+\rho\Omega_{k}^{2}$,
with $\Omega_{k}=\sqrt{\left(\kappa k^{4}+\Sigma_{k}\right)/\rho}$. Since, according to Eq.~\eqref{eq:Sigma_T_0}, the self-energy $\Sigma_{k}$
goes to a constant at $T=0$, we will ignore the temperature dependence
of the self-energy for $T\simeq0$, approximating $\pd\Omega_{k}/\pd T\simeq0$. With this approximations, the thermal expansion is given by Eq.~\eqref{eq:expansion_qh} with the replacement $\omega_{k,F}\rra\Omega_{k}$. Now the integral is finite both in the IR and in the UV and we can take $q_{\text{min}}\rra0$ and $\Lambda\rra\infty$. Since in the $T\rra0$ limit the main contribution will
come from small momentum modes, we approximate $\rho\Omega_{k}^{2}=\kappa k^{4}+\Sigma_{k}\simeq\ka k_{c}^{\eta}k^{4-\eta}$, where $\eta$ is a characteristic exponent whose value depends on the approximation used to solve the theory. After a straightforward calculation
we obtain 
\begin{equation}
\alpha_{V}=-\frac{k_{B}}{2\pi\left(4-\eta\right)\kappa}\left(\frac{2\rho^{1/2}k_{B}T}{\hbar\kappa^{1/2}k_{c}^{2}}\right)^{2\eta/\left(4-\eta\right)}I_{\eta},\label{eq:th_exp_T_0}
\end{equation}
where we have defined $I_{\eta}=\int_{0}^{\infty}dxx^{(4+\eta)/(4-\eta)}/\sinh^{2}\left(x\right)$,
which for $\eta=2$ evaluates to $I_{2}=3\zeta(3)/2$. Since we have
obtained $\eta=2$ within first order perturbation theory as well as 
in the partially self-consistent approximation, we obtain that $\alpha_{V}\propto-T^{2}$
at low temperature. Most importantly, this result correctly predicts a vanishing thermal
expansion coefficient for $T\rra0$, satisfying the third law of thermodynamics even in the limit of an infinite membrane, $L\rra\infty$.

\subsection{Specific heat}

Another important thermodynamic physical property, probing the low-energy
elementary excitations in the system, is the specific heat. We are
working in an ensemble with constant external stress $\sigma_{ij}$.
The specific heat at constant pressure/stress can be computed from 
\begin{equation}
c_{p}=\left(\frac{\pd\text{H}}{\pd T}\right)_{p},
\end{equation}
where $\text{H}=U-u_{ij}\sigma_{ij}$ is the enthalpy of the system
per unit area, with $U$ the total energy of the system per unit area.
Since we are working at zero external stress, the enthalpy coincides
with the total internal energy. The total energy of the system can
be expressed in terms of two-point correlation functions, using
a modified Migdal-Galitskii-Koltun energy sum rule \cite{Migdal_1958,Koltun_1974}.
The total energy of the membrane per unit area can be written as $U=U^{(\text{out})}+U^{\text{(in)}}$
with (see Appendix~\ref{sec:Appendix_MGK_sum} for the proof) 
\begin{eqnarray}
U^{(\text{out})} & = & \frac{1}{4}\frac{1}{\beta V}\sum_{\bm{k}}\left(3\rho\left(ik_{n}\right)^{2}+\kappa k^{4}\right)G_{\bm{k}},\label{eq:U_out}\\
U^{(\text{in})} & = & \frac{1}{\beta V}\sum_{\bm{q}}\rho\left(iq_{n}\right)^{2}D_{\bm{q}}^{ii}.\label{eq:U_in}
\end{eqnarray}
Although in the anharmonic theory
in-plane and out-of-plane modes are coupled, we can attribute $U^{(\text{out})}$
mostly to out-of-plane modes and $U^{(\text{in})}$ mostly to in-plane
modes. In the same way, the specific heat can also be split in these
two contributions, $c_{p}=c_{p}^{(\text{out})}+c_{p}^{(\text{in})}$,
where $c_{p}^{(\text{out})}=\pd U^{(\text{out})}/\pd T$ and $c_{p}^{(\text{in})}=\pd U^{(\text{in})}/\pd T$.
As previously, we ignore the frequency dependence of the 
self-energy, and obtain the energy per unit area 
\begin{equation}
U^{(\text{out})}=\frac{\hbar}{4}\int\frac{d^{2}k}{\left(2\pi\right)^{2}}\frac{\coth\left(\beta\hbar\Omega_{k}/2\right)}{2\rho\Omega_{k}}\left(3\rho\Omega_{k}^{2}+\kappa k^{4}\right).
\end{equation}
In the $T\rra0$ limit, making the same approximations as for the thermal expansion, 
we can write the specific heat at constant
pressure per unit area as 
\begin{equation}
c_{p}^{(\text{out})}=\frac{3}{8\pi}k_{B}k_{c}^{2}\left(\frac{2\rho^{1/2}k_{B}T}{\hbar\kappa^{1/2}k_{c}^{2}}\right)^{4/(4-\eta)}L_{\eta},
\end{equation}
where $L_{\eta}=\int_{0}^{\infty}dxx^{(8-\eta)/(4-\eta)}/\sinh^{2}(x)$.
For $\eta=2$, which is the value corresponding to the approximations used in this paper, we have $L_{2}=3\zeta(3)/2$ and $c_{p}^{(\text{out})}\propto T^{2}$.
This result is to be contrasted with the one obtained at the harmonic
level, which would predict $c_{p}^{(\text{out})}\propto T$. It is a consequence of the change of dispersion of flexural modes from $\sim k^2$ to $\sim k$ as discussed after Eq.\eqref{Eq:qLT}.
Regarding
the contribution mostly due to in-plane modes, we can check that interactions
lead only to a logarithmic correction of the in-plane modes correlation
function (see Appendix~\ref{sec:Appendix_self_energy}), which we
will neglect. Therefore, the contribution mostly due to the in-plane
modes reduces to the non-interacting one, which for $T\rra0$ reduces
to the expected  $T^{2}$ dependence 
\begin{equation}
c_{p}^{(\text{in})}=k_{B}\left(\frac{2k_{B}T}{\hbar}\right)^{2}\left(\frac{\rho}{\la+2\mu}+\frac{\rho}{\mu}\right)L_{2}.
\end{equation}
Therefore, taking into account at the same level anharmonic and quantum
effects, one predicts an intermediate behavior $T<T^{4/(4-\eta)}\leqslant T^{2}$
resulting from the coupling between in-plane and out-of-plane
modes. To first order in perturbation theory, both $c_{p}^{(\text{out})}$
and $c_{p}^{(\text{in})}$ are proportional to  $T^{2}$.
Notice, that the harmonic theory calculated for graphene \cite{Popov_2002,MM05}, predicts
$c_{p}\propto T$ up to temperatures as high as $100\text{ K}$ which is about our $T^*$. 
That is why we believe that the  linear $T$ dependence should not be observable  in graphene
for which we  predict instead a $T^{2}$ dependence.

\section{Conclusions}

In summary, in this paper we have calculated several thermodynamic properties of crystalline membranes in the low temperature
quantum regime. Toward that end we have employed both a first order perturbation
theory as well as a one-loop self-consistent approximation in which we have ignored any possible renormalization of the in-plane Lam\'e elastic constants. We have derived the effective action for  the out-of-plane modes by integrating out exactly
the  in-plane modes. This procedure leads to frequency dependent anharmonic interactions (retardation effects) which we have shown to be the dominant  effect in the zero temperature
limit.
This is to be contrasted with the high temperature classical regime,
where retardation can be ignored. We have further evaluated the leading  of
the anharmonic out-of-plane mode self-energy in the $T\rra0$ limit
and estimated the available phase space, described by a crossover
momentum $k_{c}$, which defines a wavelength above which anharmonic effects dominate the theory in the quantum regime. 
For graphene we estimate $k_{c}\sim 0.1$ \AA$^{-1}$, about 0.6 of the value estimated for the crossover from the harmonic to the anharmonic regime in the classical case at room temperature \cite{LF09}.
Based on this result, we estimate a crossover
temperature $T^{*}$ between the classical and quantum regimes. For typical graphene
parameters, this crossover temperature is $T^{*} \sim 70 - 90$ K.

By using the calculated correlation functions in the quantum anharmonic regime, we establish the temperature dependence of 
thermodynamic properties. In the $T\rra0$ limit,
we find a power-law behavior for both the thermal expansion coefficient $\alpha_V$ and the specific heat $c_p$.
In general they are characterized by
an anomalous exponent related to the characteristic exponent $\eta$ of
the elementary excitations, namely  $\alpha_{V}\propto T^{2\eta/(4-\eta)}$
and $c_{p}\propto T^{4/(4-\eta)}$.   To first order
perturbation theory, as well as in the one-loop self-consistent approximation we find  $\eta=2$, which means that
both $\alpha_V$ and  $c_p$  are proportional to $T^2$. 

This work
is a first step towards  the full understanding of the physics
of a quantum crystalline membrane.
We know that for a classical crystalline
membrane, it is necessary to go beyond perturbation theory and use a
more elaborate technique, such as a full self-consistent calculation. Also for the quantum case, more advanced methods 
such as quantum Monte Carlo or functional renormalization group are needed to solve the problem quantitatively.  
Nevertheless, the perturbative
calculation that we have presented is already sufficient to show that a simultaneous  treatment
of quantum and  anharmonic effects is necessary to have a vanishing thermal expansion and specific heat at zero temperature,
in accordance with the third law of thermodynamics. This approach also allows us to estimate the crossover temperature between the classical and the quantum regime.

\begin{acknowledgments}
We thank SURFsara (www.surfsara.nl) for the support Grant No. (MP-282-13) in using the Lisa Compute Cluster.
B.A. acknowledges support from
Funda\c{c}\~{a}o para a Ci\^{e}ncia e a Tecnologia (Portugal), through Grant.
No. SFRH/BD/78987/2011.  R.R. acknowledges financial support from the Juan de la Cierva Programe
(MEC, Spain).  
E.C. acknowledges support from the European
project FP7-PEOPLE-2013-CIG "LSIE\_2D" and Italian
National MIUR Prin project 20105ZZTSE.
R.R. and F.G. thank financial support from MINECO, Spain, through Grant No. FIS2011-23713. M.I.K and A.F. acknowledge 
funding from the European Union Seventh 
Framework Programme under grant agreement n604391 Graphene Flagship.
\end{acknowledgments}

\appendix
%dummy comment inserted by tex2lyx to ensure that this paragraph is not empty

\section{Derivation of effective action for the out-of-plane modes\label{sec:Appendix_eff_action}}

In this appendix, we briefly summarize the steps performed to derive
the effective action \eqref{eq:S_eff}. Notice that the cubic action
$\mathcal{S}_{\text{int}}^{(3)}\left[\vec{u},h\right]$, corresponding to Eq.~\eqref{eq:cubic_u-h}, can be written in
terms of Fourier components as 
\begin{equation}
\mathcal{S}_{\text{int}}^{(3)}\left[\vec{u},h\right]=\frac{i}{2\sqrt{\beta V}}\sum_{\bm{k},\bm{q}}u_{\bm{q}}^{i}c^{ijkl}q_{j}\left(k-q\right)_{k}k_{l}h_{\bm{k}-\boldsymbol{q}}h_{-\bm{k}},
\end{equation}
where we have introduce the elastic moduli tensor for an isotropic membrane
$c^{ijkl}=\la\de^{ij}\de^{kl}+\mu\left(\de^{ik}\de^{jl}+\de^{il}\de^{jk}\right).$
Integrating out the field $\vec{u}$, amounts to performing a Gaussian
integration of the form $\int dxe^{-\frac{1}{2a}x^{2}-bx}=e^{\frac{1}{2}b^{2}a}\left(\int dxe^{-\frac{1}{2a}x^{2}}\right)$.
As a results, the partition function can be written as 
\begin{align}
Z & =\int D\left[\vec{u},h\right]\exp\left(-\mathcal{S}\left[\vec{u},h\right]\right)\nonumber \\
 & =Z_{0}\left[\vec{u}\right]\int D\left[h\right]\exp\left(-\mathcal{S}_{\text{eff}}\left[h\right]\right),
\end{align}
where $Z_{0}\left[\vec{u}\right]=\int D\left[\vec{u}\right]\exp\left(-\int_{0}^{\beta}d\tau\int d^{2}x\mathcal{L}_{u}^{0}\left[\vec{u}\right]\right)$
is the non-interacting partition function for the in-plane modes.
Integrating out the in-plane modes will therefore generate a new
quartic interaction term for the out-of-plane modes that is mediated
by the in-plane modes. Therefore, we obtain an effective Euclidean
action of the form of Eq.~\eqref{eq:S_eff}, with the interaction tensor
given by 
\begin{equation}
R_{\bm{q}}^{ijkl}=c^{ijkl}-c^{iji^{\prime}j^{\prime}}\left\langle u_{i^{\prime}j^{\prime},\bm{q}}u_{k^{\prime}l^{\prime},\bm{-q}}\right\rangle _{0}c^{k^{\prime}l^{\prime}kl},\label{eq:R_ijkl}
\end{equation}
where $\left\langle \right\rangle _{0}$ represents averaging with respect 
to the harmonic theory and $u_{ij}=\left(\pd_{i}u_{j}+\pd_{j}u_{i}\right)/2$
is the in-plane strain tensor. The first term of $R_{\bm{q}}^{ijkl}$
is due to the quartic interaction $\mathcal{L}_{\text{int}}^{(4)}\left[h\right]$, as given by Eq. \eqref{eq:quartic_on-site}, while the second term is the in-plane mode mediated interaction due to the cubic term $\mathcal{L}_{\text{int}}^{(3)}\left[\vec{u},h\right]$, Eq. \eqref{eq:cubic_u-h}.
From Eq.~\eqref{eq:R_ijkl}, it is easy to see that $R_{\bm{q}}^{ijkl}$
obeys the same symmetries as the elastic moduli tensor $c^{ijkl}$, i.e.,
$R_{\bm{q}}^{ijkl}=R_{\bm{q}}^{jikl}=R_{\bm{q}}^{klij}$. Just like
in the classical problem, it is necessary to analyse the cases $\bm{q}\neq0$
and $\bm{q}=0$ separately \cite{NPW04}. The in-plane strain tensor
$u_{ij}$ must be split into its $\bm{q}=0$ homogeneous component,
$u_{ij}^{0}$, and $\bm{q}\neq0$ components, which can be expressed
in terms of phonon modes. For $\bm{q}\neq0$, we have 
\begin{align}
\left\langle u_{ij,\bm{q}}u_{kl,\bm{-q}}\right\rangle _{0} & =\frac{1}{2}q^{2}D_{\bm{q}}^{L,0}\left(P_{ik}^{L}P_{jl}^{L}+P_{jk}^{L}P_{il}^{L}\right)\nonumber \\
 & +\frac{1}{2}q^{2}D_{\bm{q}}^{T,0}\left(P_{ik}^{L}P_{jl}^{T}+P_{jk}^{L}P_{il}^{T}\right)
\end{align}
where $P_{ij}^{L}=q_{i}q_{j}/q^{2}$ and $P_{ij}^{T}=\de_{ij}-q_{i}q_{j}/q^{2}$
are, respectively, the longitudinal and transverse projectors along
the vector $\vec{q}$. Therefore, for $\bm{q}\neq0$, the in-plane mode
mediated interaction can be written as  
\begin{align}
 & c^{iji^{\prime}j^{\prime}}\left\langle u_{i^{\prime}j^{\prime},\bm{q}}u_{k^{\prime}l^{\prime},\bm{-q}}\right\rangle _{0}c^{k^{\prime}l^{\prime}kl}=\nonumber \\
= & \la^{2}q^{2}D_{\bm{q}}^{L,0}P_{ij}^{T}P_{kl}^{T}+\left(\la+2\mu\right)^{2}q^{2}D_{\bm{q}}^{L,0}P_{ij}^{L}P_{kl}^{L}\nonumber \\
+ & \la\left(\la+2\mu\right)q^{2}D_{\bm{q}}^{L,0}\left(P_{ij}^{L}P_{kl}^{T}+P_{ij}^{T}P_{kl}^{L}\right)\nonumber \\
+ & \mu^{2}q^{2}D_{\bm{q}}^{T,0}\left(P_{ik}^{T}P_{jl}^{L}+P_{il}^{T}P_{jk}^{L}+P_{jk}^{T}P_{il}^{L}+P_{ik}^{L}P_{jl}^{T}\right).
\end{align}
The elastic moduli tensor can also be decomposed in terms of longitudinal
and transverse projectors 
\begin{align}
c^{ijkl} & =\la P_{ij}^{T}P_{kl}^{T}+\mu\left(P_{ik}^{T}P_{jl}^{T}+P_{il}^{T}P_{jk}^{T}\right)\nonumber \\
 & +\left(\la+2\mu\right)P_{ij}^{L}P_{kl}^{L}+\la\left(P_{ij}^{T}P_{kl}^{L}+P_{ij}^{L}P_{kl}^{T}\right)\nonumber \\
 & +\mu\left(P_{ik}^{T}P_{jl}^{L}+P_{ik}^{L}P_{jl}^{T}+P_{il}^{T}P_{jk}^{L}+P_{il}^{L}P_{jk}^{T}\right).
\end{align}
In 2D, we have  $P_{ij}^{T}P_{kl}^{T}=P_{ik}^{T}P_{jl}^{T}=P_{il}^{T}P_{jk}^{T}$.
Therefore, in 2D $R_{\bm{q}}^{ijkl}$ has only 4 independent
components. As a result,  the effective interaction for $\bm{q}\neq0$ can be
expressed in the basis $\left\{ \hat{e}_{\parallel},\hat{e}_{\perp}\right\} $
as 
\begin{align}
R_{\bm{q}}^{ijkl} & =R_{\bm{q}}^{\perp\perp\perp\perp}\hat{e}_{\perp}^{i}\hat{e}_{\perp}^{j}\hat{e}_{\perp}^{k}\hat{e}_{\perp}^{l}+R_{\bm{q}}^{\parallel\parallel\parallel\parallel}\hat{e}_{\parallel}^{i}\hat{e}_{\parallel}^{j}\hat{e}_{\parallel}^{k}\hat{e}_{\parallel}^{l}\nonumber \\
 & +R_{\bm{q}}^{\perp\perp\parallel\parallel}\left(\hat{e}_{\perp}^{i}\hat{e}_{\perp}^{j}\hat{e}_{\parallel}^{k}\hat{e}_{\parallel}^{l}+(ij\leftrightarrow jl)\right)\nonumber \\
 & +R_{\bm{q}}^{\perp\parallel\perp\parallel}\left(\hat{e}_{\perp}^{i}\hat{e}_{\parallel}^{j}\hat{e}_{\perp}^{k}\hat{e}_{\parallel}^{l}+(i\leftrightarrow j)+(k\leftrightarrow l)+(ij\leftrightarrow jl)\right),
\end{align}
with $R_{\bm{q}}^{\perp\perp\perp\perp}$, $R_{\bm{q}}^{\parallel\parallel\parallel\parallel}$,
$R_{\bm{q}}^{\perp\perp\parallel\parallel}$ and $R_{\bm{q}}^{\perp\parallel\perp\parallel}$
given by Eqs.~\eqref{eq:R_tttt}-\eqref{eq:R_tptp}. Notice that for
the generalized problem of a $D>2$ dimensional membrane, $R_{\bm{q}}^{ijkl}$
will have an extra independent component which involves only the shear
modulus $\mu$. For the $\bm{q}=0$ component, the quadratic Lagrangian
density \eqref{eq:harmonic_in} reads $\mathcal{L}_{u}^{0}\left[u_{ij}^{0}\right]=\frac{1}{2}c_{ijkl}u_{ij}^{0}u_{kl}^{0}$.
Therefore, $\left\langle u_{ij}^{0}u_{kl}^{0}\right\rangle _{0}$
is simply the  tensor $c_{ijkl}^{-1}$, which for an isotropic
membrane reads $c_{ijkl}^{-1}=-\frac{\la}{4\mu\left(\la+\mu\right)}\de_{ij}\de_{kl}+\frac{1}{4\mu}\left(\de_{ik}\de_{jl}+\de_{il}\de_{jk}\right)$.
For $\bm{q}=0$, we have  $R_{\bm{q}=0}^{ijkl}=c^{ijkl}-c^{iji^{\prime}j^{\prime}}c_{i^{\prime}j^{\prime}k^{\prime}l^{\prime}}^{-1}c^{k^{\prime}l^{\prime}kl}=0$.
This justifies the exclusion of the $\bm{q}=0$ component from
the interaction term in Eq.~\eqref{eq:S_eff}, just as in the classical
problem.

\section{Perturbative evaluation of the self-energies \label{sec:Appendix_self_energy}}

\subsection{Out-of-plane mode self-energy\label{sec:Appendix_self_energy_flexural}}

To first order in the interaction, the self-energy of the out-of-plane mode is given by 
\begin{equation}
\Sigma_{\boldsymbol{k}}=\frac{1}{\beta V}\sum_{\boldsymbol{q}}R_{\bm{q}}^{ijkl}\left(k+q\right)_{i}k_{j}\left(k+q\right)_{k}k_{l}G_{\boldsymbol{k+q}}^{0},
\end{equation}
which can be decomposed into the sum of one {\it classical} term plus 4 {\it quantum}
terms 
\begin{equation}
\Sigma_{\boldsymbol{k}}=\Sigma_{\boldsymbol{k}}^{\text{cl}}+\Sigma_{\boldsymbol{k}}^{\perp\perp\perp\perp,\text{qt}}+\Sigma_{\boldsymbol{k}}^{\parallel\parallel\parallel\parallel,\text{qt}}+\Sigma_{\boldsymbol{k}}^{\perp\perp\parallel\parallel,\text{qt}}+\Sigma_{\boldsymbol{k}}^{\perp\parallel\perp\parallel,\text{qt}},
\end{equation}
where the {\it classical} contribution is given by 
\begin{equation}
\Sigma_{\boldsymbol{k}}^{\text{cl}}=\frac{4\mu\left(\la+\mu\right)}{\la+2\mu}\frac{1}{\beta V}\sum_{\bm{q}}g_{\vec{k},-\vec{k},\vec{q}}^{\perp\perp\perp\perp}G_{\boldsymbol{k+q}}^{0}
\end{equation}
and the {\it quantum} terms have the general form 
\begin{equation}
\Sigma_{\boldsymbol{k}}^{M,\text{qt}}=C^{M}\frac{1}{\beta V}\sum_{\bm{q}}g_{\vec{k},-\vec{k},\vec{q}}^{M}\frac{-\rho\left(iq_{n}\right)^{2}}{-\rho\left(iq_{n}\right)^{2}+\rho\omega_{q,M}^{2}}G_{\boldsymbol{k+q}}^{0},
\end{equation}
with the label $M$ running over$\left\{ ^{\perp\perp\perp\perp},^{\parallel\parallel\parallel\parallel},^{\perp\perp\parallel\parallel},^{\perp\parallel\perp\parallel}\right\} $.
$g_{\vec{k},\vec{p},\vec{q}}^{M}$ are given by Eqs.~\eqref{eq:g_tttt}-\eqref{eq:g_tptp}; $C^{M}$ are given by $C^{\perp\perp\perp\perp}=\la^{2}/\left(\la+2\mu\right)$,
$C^{\parallel\parallel\parallel\parallel}=\mbox{\ensuremath{\la}}+2\mu$,
$C^{\perp\perp\parallel\parallel}=\la$, $C^{\perp\parallel\perp\parallel}=\mu$; and $\omega_{q,M}=\omega_{q,L}$ for , $M=^{\perp\perp\perp\perp},^{\parallel\parallel\parallel\parallel},^{\perp\perp\parallel\parallel}$ while $\omega_{q,\perp\parallel\perp\parallel}=\omega_{q,T}$.

\subsubsection{Contribution from classical term: $\Sigma_{\boldsymbol{k}}^{\text{cl}}$}

Performing the sum over Matsubara frequencies for the {\it classical} contribution
one obtains 
\begin{equation}
\Sigma_{k}^{\text{cl}}=\frac{4\mu\left(\la+\mu\right)}{\la+2\mu}\hbar\int\frac{d^{2}q}{\left(2\pi\right)^{2}}k^{4}\sin^{4}\theta\frac{\coth\left(\beta\hbar\omega_{k+q,F}/2\right)}{2\rho\omega_{k+q,F}}.
\end{equation}
In the high-temperature limit, $T\rra\infty$, we have $\coth\left(\beta\hbar\omega/2\right)\simeq2k_{B}T/(\hbar\omega)$,
and we recover the well known result \cite{NPW04} 
\begin{equation}
\lim_{T\rra\infty}\Sigma_{\boldsymbol{k}}^{\text{cl}}\simeq\frac{4\mu\left(\la+\mu\right)}{\left(\la+2\mu\right)\kappa}\frac{3k_{B}T}{16\pi}k^{2}.
\end{equation}
In the zero temperature limit, $T\rra0$, we have $\coth\left(\beta\hbar\omega/2\right)\simeq1$.
Therefore, the classical contribution becomes 
\begin{equation}
\lim_{T\rra0}\Sigma_{k}^{\text{cl}}\simeq\frac{4\mu\left(\la+\mu\right)}{\la+2\mu}\frac{\hbar}{2\sqrt{\rho\kappa}}\frac{3}{16\pi}k^{4}\log\left(\frac{\Lambda}{k}\right),
\end{equation}
where $\Lambda$ is a UV momentum cutoff, which we identify as the
Debye momentum. This {\it classical} contribution has to be compared with
the {\it quantum} ones.

\subsubsection{Contribution from quantum terms: $\Sigma_{\boldsymbol{k}}^{\perp\perp\perp\perp,\text{qt}}$,
$\Sigma_{\boldsymbol{k}}^{\parallel\parallel\parallel\parallel,\text{qt}}$,
$\Sigma_{\boldsymbol{k}}^{\perp\perp\parallel\parallel,\text{qt}}$
and $\Sigma_{\boldsymbol{k}}^{\perp\parallel\perp\parallel,\text{qt}}$}

Performing the sum over Matsubara frequencies for the {\it quantum} contributions
one obtains [with a small  change of notation $\Sigma_{\bm{k}}=\Sigma_{k}(ik_{n})$]
\begin{widetext}
\begin{align}
\Sigma_{k}^{M,\text{qt}}(ik_{n}) & =\frac{\pv}{V}\sum_{\vec{q}}g_{\vec{k},-\vec{k},\vec{q}}^{M}\int\frac{dy}{\pi}b(y)\im R_{q}^{M,\text{qt}}\left(y+i0^{+},q\right)G_{k+q}^{0}(ik_{n}+y)\nonumber \\
 & +\frac{\pv}{V}\sum_{\vec{q}}g_{\vec{k},-\vec{k},\vec{q}}^{M}\int\frac{dy}{\pi}b(y)R_{q}^{M,\text{qt}}(y-ik_{n},q)\im G_{k+q}^{0}(y+i0^{+}),
\end{align}
\end{widetext}where $\pv$ denotes Cauchy principal value. Performing
the analytic continuation $ik_{n}\rra\om+i0^{+}$ and taking the real
part we obtain 
\begin{equation}
\re\Sigma_{k}^{M,\text{qt}}(\omega)=\frac{C^{M}}{\rho}\frac{\pv}{V}\sum_{\vec{q}}g_{\vec{k},-\vec{k},\vec{q}}^{M}\mathcal{K}^{M,\text{qt}}(\omega,k,q),
\end{equation}
where  we have defined
\begin{align}
\mathcal{K}^{M,\text{qt}}(\omega,k,q) & =-\frac{\hbar b(\omega_{q,M})}{2\omega_{q,(A)}}\frac{\omega_{q,M}^{2}}{-\left(\omega+\omega_{q,M}\right)^{2}+\omega_{k+q,F}^{2}}\nonumber \\
 & +\frac{\hbar b(-\omega_{q,M})}{2\omega_{q,(A)}}\frac{\omega_{q,M}^{2}}{-\left(\omega-\omega_{q,M}\right)^{2}+\omega_{k+q,F}^{2}}\nonumber \\
 & +\frac{\hbar b(\omega_{k+q,F})}{2\omega_{k+q,F}}\frac{-\left(\omega_{k+q,F}-\om\right)^{2}}{-\left(\omega_{k+q,F}-\omega\right)^{2}+\omega_{q,M}^{2}}\nonumber \\
 & -\frac{\hbar b(-\omega_{k+q,F})}{2\omega_{k+q,F}}\frac{-\left(-\omega_{k+q,F}-\om\right)^{2}}{-\left(-\omega_{k+q,F}-\omega\right)^{2}+\omega_{q,M}^{2}},
\end{align}
with $b(\omega)=\left(\exp(\beta\hbar\omega)-1\right)^{-1}$ the Bose-Einstein
distribution function. In the high temperature limit, we have $b(\omega)\simeq k_{B}T/(\hbar\omega)$
and we obtain \begin{widetext} 
\begin{equation}
\mathcal{K}_{T\rra\infty}^{M,\text{qt}}(\omega,k,q)=\frac{k_{B}T}{\omega_{k+q,F}^{2}}\frac{\omega^{2}\left(\omega^{2}-\omega_{q,M}^{2}-\omega_{k+q,F}^{2}\right)}{\left[\left(\omega_{k+q,F}+\omega_{q,M}\right)^{2}-\omega^{2}\right]\left[\left(\omega_{k+q,F}-\omega_{q,M}\right)^{2}-\omega^{2}\right]}.
\end{equation}
\end{widetext} In this form we can see explicitly that for $\omega=0$
we have $\mathcal{K}_{T\rra\infty}^{M,\text{qt}}(0,k,q)=0$, and the
{\it quantum} terms do not give any contribution to the self-energy in the
$T\rra\infty$ limit. Notice, however, that even if we take $T\rra\infty$
but keep $\omega\neq0$, we obtain a non zero value of $\re\Sigma_{k}^{M,\text{qt}}(\omega)$.
The analysis of this situation is subtle. One can check that for finite
$\omega$, the integrations over momentum involved in computing $\Sigma_{\boldsymbol{k}}^{\parallel\parallel\parallel\parallel,\text{qt}}$
and $\Sigma_{\bm{k}}^{\parallel\perp\parallel\perp,\text{qt}}$ are
logarithmically divergent due to the point $\vec{q}=-\vec{k}$. This is a pathology of the first order
perturbation theory that should disappear if a more complete self-consistent
calculation is performed. If we replace
the dispersion relation of the out-of-plane modes $\omega_{k,F}\rra\Omega_{k}\propto k^{(4-\eta)/2}$,
for any $\eta>0$, the integrals become finite. Assuming that such regularization is performed, and to lowest order in the frequency, the quantum contributions
are suppressed by a factor of $\omega^{2}/(c_{L/T}k)^{2}\log\left(c_{L/T}k/\omega\right)$
when compared with the contribution from $\Sigma_{k}^{\text{cl}}$
in the $T\rra\infty$ limit. We remind the reader that the dispersion of the
physical excitation is obtained, to first order in perturbation theory,
by $\Omega_{k}^{2}=\omega_{k,F}^{2}+\re\Sigma_{k}(\omega_{k,F})/\rho$.
Therefore, for the relevant long wavelength limit, $k\rra0$, 
the quantum contributions can be ignored in the $T\rra\infty$ limit, as
expected. However, this situation changes dramatically in the quantum,
$T\rra0$, limit. In this limit, we have $b(\omega)\simeq-\Theta(-\omega)$, where $\Theta(x)$ is the step function,
and therefore we obtain 
\begin{widetext} 
\begin{equation}
\mathcal{K}_{T\rra0}^{M,\text{qt}}(\omega,k,q)=\frac{\hbar}{2\omega_{k+q,F}}\frac{\omega_{k+q,F}\left(\omega_{k+q,F}+\omega_{q,M}\right)-\omega^{2}}{\left(\omega_{k+q,F}+\omega_{q,M}\right)^{2}-\omega^{2}}.
\end{equation}
\end{widetext} We can see that we obtain finite contributions
even if we set $\omega=0$, in which case we have 
\begin{equation}
\mathcal{K}_{T\rra0}^{M,\text{qt}}(0,k,q)=\frac{\hbar}{2}\frac{1}{\omega_{k+q,F}+\omega_{q,M}}.
\end{equation}
With this in mind we will focus on the static, $\omega=0$, behavior of the self-energy, which will be the dominant
one in the long wavelength limit, $k\rra0$. Therefore, we set $\omega=0$ and
expand to lowest order in $k$. The different {\it quantum} contributions
to the self-energy yield 
\begin{align}
\Sigma_{k,T\rra0}^{\perp\perp\perp\perp,\text{qt}}(0) & \simeq\frac{\hbar}{2\sqrt{\rho\kappa}}\frac{\la^{2}}{\la+2\mu}\frac{3k^{4}}{16\pi}\log\left(1+\frac{\Lambda}{q_{L}}\right),\label{eq:SelfT0tttt}\\
\Sigma_{k,T\rra0}^{\parallel\parallel\parallel\parallel}(0) & \simeq\frac{\hbar}{2\sqrt{\rho\kappa}}\frac{(\la+2\mu)^{2}}{\kappa}\frac{k^{2}}{4\pi}f\left(\frac{\Lambda}{q_{L}}\right),\\
\Sigma_{k,T\rra0}^{\perp\perp\parallel\parallel}(0) & \simeq\frac{\hbar}{2\sqrt{\rho\kappa}}\frac{\la k^{4}}{8\pi}\left(\frac{2\Lambda}{q_{L}+\Lambda}-\log\left(1+\frac{\Lambda}{q_{L}}\right)\right),\\
\Sigma_{k,T\rra0}^{\perp\parallel\perp\parallel}(0) & \simeq\frac{\hbar}{2\sqrt{\rho\kappa}}\frac{\mu^{2}}{\kappa}\frac{k^{2}}{4\pi}f\left(\frac{\Lambda}{q_{T}}\right).\label{eq:SelT0tptp}
\end{align}
where we have once again imposed an UV momentum cutoff $\Lambda$,
with $f\left(x\right)$ and $q_{L/T}$ as given by Eqs. \eqref{Eq:f} and \eqref{Eq:qLT}, respectively.
Therefore, for $T\rra0$, the main contribution in the long wavelength
limit comes from $\Sigma_{k}^{\parallel\parallel\parallel\parallel}(0)$
and $\Sigma_{k}^{\parallel\perp\parallel\perp}(0)$. 

For the on-shell case $\omega=\omega_{k,F}$ and in the long wavelength
limit $k\rra0$, the results \eqref{eq:SelfT0tttt}-\eqref{eq:SelT0tptp}
are not changed. This tells us that, at least at the perturbative level, the frequency
dependence of the self-energy can be neglected for physical excitations in the long wavelength limit.

\subsection{In-plane mode self-energy\label{sec:Appendix_self_energy_inplane}}

Notice, that although $\vec{u}$ no longer appears in $\mathcal{S}_{\text{eff}}\left[h\right]$,
that does not mean that it is not affected by the interactions. When
computing any correlation function, one must remember that in the
process of integrating out the in-plane modes, they were shifted by
\begin{equation}
u_{\bm{q}}^{i}\rra v_{\boldsymbol{q}}^{i}+\frac{i}{2\sqrt{\beta V}}\sum_{\bm{k},\bm{q}}\left(D_{\bm{q}}^{0}\right)^{ij}c^{jklm}q_{k}\left(k+q\right)_{l}k_{m}h_{\bm{k}+\boldsymbol{q}}h_{-\bm{k}},
\end{equation}
where the field $v_{\bm{q}}^{i}$ is a free field, with Lagrangian
given by the in-plane harmonic one, $\mathcal{L}_{u}^{0}\left[\vec{u}\right]$, as given by
Eq. \eqref{eq:harmonic_in}. Therefore, the full in-plane correlation function
is given by\begin{widetext} 
\begin{equation}
D_{\bm{q}}^{ij}=\left(D_{\bm{q}}^{0}\right)^{ij}+\frac{1}{4\beta V}\left(D_{\bm{q}}^{0}\right)^{ik}c^{klmn}q_{l}\sum_{\bm{k},\bm{p},\bm{q}}\left(k+q\right)_{m}k_{n}\left(p-q\right)_{m^{\prime}}p_{n^{\prime}}\left\langle h_{\bm{k}+\boldsymbol{q}}h_{-\bm{k}}h_{\bm{p}-\boldsymbol{q}}h_{-\bm{p}}\right\rangle c^{k^{\prime}l^{\prime}m^{\prime}n^{\prime}}q_{l^{\prime}}\left(D_{\bm{q}}^{0}\right)^{k^{\prime}j}.
\end{equation}
\end{widetext} To first order in perturbation theory, we decouple
the four-point correlation function and obtain 
\begin{equation}
D_{\bm{q}}^{ij}=\left(D_{\bm{q}}^{0}\right)^{ij}+\left(D_{\bm{q}}^{0}\right)^{ik}q_{l}c^{klmn}\Pi_{\bm{q}}^{mnm^{\prime}n^{\prime}}c^{m^{\prime}n^{\prime}k^{\prime}l^{\prime}}q_{k^{\prime}}\left(D_{-\bm{q}}^{0}\right)^{l^{\prime}j},
\end{equation}
where 
\begin{equation}
\Pi_{\boldsymbol{q}}^{ijkl}=\frac{1}{2\beta V}\sum_{\boldsymbol{k}}\left(k+q\right)_{i}k_{j}\left(k+q\right)_{k}k_{l}G_{\boldsymbol{k}}G_{\boldsymbol{k+q}}.
\end{equation}
Just like in the harmonic theory, isotropy allows us to split $D_{\bm{q}}^{ij}$
in a longitudinal and a transverse component, $D_{\bm{q}}^{ij}=D_{\bm{q}}^{L}P_{ij}^{L}+D_{\bm{q}}^{T}P_{ij}^{T}$,
where $P_{ij}^{L}=q_{i}q_{j}/q^{2}$ and $P_{ij}^{T}=\de_{ij}-q_{i}q_{j}/q^{2}$ are
the longitudinal and transverse projectors, respectively. To lowest order in perturbation
theory, the self-energies for the in-plane modes are given by 
\begin{align}
\mathcal{P}_{\bm{q}}^{L} =& -q^{2}\la^{2}\Pi_{\bm{q}}^{\perp\perp\perp\perp}-q^{2}(\la+2\mu)^{2}\Pi_{\bm{q}}^{\parallel\parallel\parallel\parallel}\nonumber \\
 & -q^{2}\la(\la+2\mu)\Pi_{\bm{q}}^{\perp\perp\parallel\parallel},\label{eq:SelfEnergy_L}\\
\mathcal{P}_{\bm{q}}^{T} =& -q^{2}\mu^{2}\Pi_{\bm{q}}^{\perp\parallel\perp\parallel},\label{eq:SelfEnerg_T}
\end{align}
where 
\begin{align}
\Pi_{\bm{q}}^{M} & =\frac{1}{2\beta V}\sum_{\boldsymbol{k}}g_{k,-k,q}^{M}G_{\boldsymbol{k}}^{0}G_{\boldsymbol{k+q}}^{0}.
\end{align}
Performing the sum over Matsubara frequencies, making the analytic
continuation, $iq_{n}\rra\omega+i0^{+}$, and taking the real part, we
obtain 
\begin{equation}
\re\Pi_{q}^{M}(\omega)=\frac{\pv}{V}\sum_{\vec{k}}g_{\vec{k},-\vec{k},\vec{q}}^{M}\mathcal{F}(\omega,q,k),
\end{equation}
with 
\begin{align}
\mathcal{F}(\omega,q,k) & =\frac{\hbar}{2\rho^{2}\omega_{k,F}}\frac{b(\omega_{k,F})}{-\left(\omega_{k,F}+\omega\right)^{2}+\omega_{k+q,F}^{2}}\nonumber \\
 & -\frac{\hbar}{2\rho^{2}\omega_{k,F}}\frac{b(-\omega_{k,F})}{-\left(-\omega_{k,F}+\omega\right)^{2}+\omega_{k+q,F}^{2}}\nonumber \\
 & +\frac{\hbar}{2\rho^{2}\omega_{k+q,F}}\frac{b(\omega_{k+q,F})}{-\left(\omega_{k+q,F}-\omega\right)^{2}+\omega_{k,F}^{2}}\nonumber \\
 & -\frac{\hbar}{2\rho^{2}\omega_{k+q,F}}\frac{b(-\omega_{k+q,F})}{-\left(-\omega_{k+q,F}-\omega\right)^{2}+\omega_{k,F}^{2}}.
\end{align}
We focus on the $T\rra0$ limit, where $\mathcal{F}(\omega,q,k)$
simplifies to 
\begin{equation}
\mathcal{F}_{T\rra0}(\omega,q,k)=\frac{\hbar}{2\rho^{2}}\frac{\omega_{k,F}+\omega_{k+q,F}}{\omega_{k,F}\omega_{k+q,F}\left[\left(\omega_{k,F}+\omega_{k+q,F}\right)^{2}-\omega^{2}\right]},
\end{equation}
Focusing on the case with $\omega\rra0$ and $q\rra0$, the integration
over momentum is effectively cutoff at small momenta by the largest of these quantities. Therefore,
apart from a numerical factor coming from the angular integration, we
obtain 
\begin{equation}
\lim_{T\rra0}\re\Pi_{q}^{M}(\omega)\propto\frac{\hbar}{2\rho^{1/2}\kappa^{3/2}}\log\left(\frac{\Lambda^{2}}{\max\left(\sqrt{\kappa/\rho}\omega,q^{2}\right)}\right).
\end{equation}
Therefore, we will just have a weak logarithmic correction to the
correlation function of the in-plane modes. Notice that the minus
sign in Eqs.~\eqref{eq:SelfEnerg_T} and \eqref{eq:SelfEnergy_L} leads
to a reduction of the in-plane elastic constants. To first order in
perturbation theory, the in-plane mode dispersion relations would
be modified to $\omega_{q,L/T}^{2}\rra\omega_{q,L/T}^{2}+\mathcal{P}^{T/L}\left(\omega_{q,L/T}\right)/\rho$.
Taking the limit $q\rra0$, we would obtain a negative dispersion
relation, indicating that the theory is unstable. We attribute this,
not to a physical instability of the membrane, but to a breakdown of
the perturbation theory, showing that one should go beyond the first order.

\section{\label{sec:Appendix_MGK_sum}Migdal-Galitskii-Koltun energy sum}

In this appendix, we will prove Eqs.~\eqref{eq:U_out} and \eqref{eq:U_in}.
The Migdal-Galitskii-Koltun energy sum \cite{Migdal_1958,Koltun_1974} allows
one to express the total energy of a system with quartic interactions
just in terms of two-point correlation functions. In the following,
we will prove a similar result but for the case of a crystalline membrane,
which contains both quartic and cubic interactions. In the canonical
quantization formalism, the Hamiltonian for a crystalline membrane
is given by\begin{widetext} 
\begin{eqnarray}
H & = & \int d^{2}x\left[\frac{1}{2\rho}\left(\pi_{h}^{2}+\vec{\pi}_{u}^{2}\right)+\frac{1}{2}\left(\ka\left(\pd^{2}h\right)^{2}+c^{ijkl}\pd_{i}u_{j}\pd_{k}u_{l}\right)\right]\nonumber \\
 &  & +\int d^{2}x\left[\frac{1}{2}c^{ijkl}\pd_{i}u_{j}\left(\pd_{k}h\pd_{l}h\right)+\frac{1}{8}c^{ijkl}\left(\pd_{i}h\pd_{j}h\right)\left(\pd_{k}h\pd_{l}h\right)\right],\label{eq:Hamiltonian}
\end{eqnarray}
\end{widetext} where $\pi_{h}$ and $\vec{\pi}_{u}$ are, respectively,
the canonical conjugate momenta of $h$ and $\vec{u}$, which obey
the equal time commutation relations $\left[h(x),\pi_{h}(x^{\prime})\right]=i\hbar\de^{(2)}\left(x-x^{\prime}\right)$
and $\left[u^{i}(x),\pi_{u}^{j}(x^{\prime})\right]=i\hbar\de^{ij}\de^{(2)}\left(x-x^{\prime}\right)$.
The proof is based on the Heisenberg equation of motion for the operators,
and the crucial point for the proof is that the Hamiltonian \eqref{eq:Hamiltonian}
has a quartic interaction for the $h$ field and a cubic interaction
involving $h$ and $\vec{u}$, such that $\vec{u}$ appears only once
in the cubic interaction. In other words, $\vec{u}$ is an interaction
mediating field. To keep the notation simple and since the essential
of the proof is not altered, instead of working with Hamiltonian \eqref{eq:Hamiltonian},
we use the Hamiltonian 
\begin{equation}
H=\frac{p_{1}^{2}}{2m_{1}}+\frac{1}{2}k_{1}x_{1}^{2}+\frac{p_{2}^{2}}{2m_{2}}+\frac{1}{2}k_{2}x_{2}^{2}+\frac{g}{2}x_{1}^{2}x_{2}+\frac{w}{8}x_{1}^{4},\label{eq:H_toy}
\end{equation}
with $p_{a}$ the canonical conjugate momentum of $x_{a}$, obeying
the equal time commutation relations $\left[x_{a},p_{b}\right]=i\hbar\de_{ab}$
($a=1,2$). Notice that the Hamiltonian \eqref{eq:H_toy} has the
same structure as \eqref{eq:Hamiltonian} if we replace $x_{1}\leftrightarrow h$
and $x_{2}\leftrightarrow\vec{u}$. We wish to evaluate the expectation
value of the energy $\left\langle H\right\rangle =T_{1}+V_{1}+T_{2}+V_{2}+W_{3}+W_{4}$,
where we have the kinetic energy of the fields, $T_{a}=\left\langle p_{a}^{2}\right\rangle /\left(2m_{a}\right)$,
the potential energy $V_{a}=k_{a}\left\langle x_{a}^{2}\right\rangle /2$,
the interaction energy due to the cubic interaction $W_{3}=g\left\langle x_{2}x_{1}^{2}\right\rangle /2$
and the interaction energy due to the quartic interaction $W_{4}=w\left\langle x_{1}^{4}\right\rangle /8$.
In the imaginary time formalism, operators evolve according to the
Heisenberg equation $\pd O(\tau)/\pd\tau=\left[H,O(\tau)\right]$.
The Heisenberg equations for the operators read
\begin{align}
i\pd_{\tau}x_{1} & =\frac{p_{1}}{m_{1}},\label{eq:x_p_1}\\
-i\pd_{\tau}p_{1} & =k_{1}x_{1}+gx_{1}x_{2}+\frac{w}{2}x_{1}^{3},\\
i\pd_{\tau}x_{2} & =\frac{p_{2}}{m_{2}},\label{eq:x_p_2}\\
-i\pd_{\tau}p_{2} & =k_{2}x_{2}+\frac{g}{2}x_{1}^{2},
\end{align}
from which the second order equation for $x_{1}$ and $x_{2}$ can
be obtained
\begin{align}
m_{1}\pd_{\tau}^{2}x_{1} & =k_{1}x_{1}+gx_{1}x_{2}+\frac{w}{2}x_{1}^{3},\label{eq:EoM2nd_x1}\\
m_{2}\pd_{\tau}^{2}x_{2} & =k_{2}x_{2}+\frac{g}{2}x_{1}^{2}.\label{eq:EoM2nd_x2}
\end{align}
Now let us define the time ordered Green's functions (recall that
a time ordered Green's function in the canonical quantization formalism,
corresponds to a correlation function in the path integral formalism)
\begin{eqnarray}
G_{ab}(\tau) & = & \left\langle T_{\tau}x_{a}(\tau)x_{b}(0)\right\rangle ,
\end{eqnarray}
where $T_{\tau}$ is the time ordering operator in imaginary time.
Using the exact eigenbasis of the interacting Hamiltonian, $H\left|n\right\rangle =E_{n}\left|n\right\rangle $,
a correlation function of the form $C_{AB}(\tau)=\left\langle T_{\tau}A(\tau)B(0)\right\rangle $,
after a Fourier transform in $\tau$, has the following Lehmann representation
\begin{align*}
C_{AB}(i\omega_{n}) & =\int_{0}^{\beta}e^{i\omega_{n}\tau}\left\langle T_{\tau}A(\tau)B(0)\right\rangle \\
 & =\frac{1}{Z}\sum_{n,m}\frac{e^{-\beta E_{m}}-e^{-\beta E_{n}}}{i\omega_{n}+E_{n}-E_{m}}A_{nm}B_{mn},
\end{align*}
where $A_{nm}=\left\langle n\right|A\left|m\right\rangle $, $Z=\sum_{n}e^{-\beta E_{n}}$
and $\omega_{n}=2\pi\beta n$ ($n\in\mathbb{Z}$) are bosonic Matsubara
frequencies. Now let us study the quantity 
\begin{equation}
\sigma_{AB}^{(2)}=\frac{1}{\beta}\sum_{i\omega_{n}}\left(i\omega_{n}\right)^{2}C_{AB}(i\omega_{n}),
\end{equation}
(a factor of $e^{i\omega_{n}\eta}$, with $\eta\rra0^{+}$, should be
added to this expression for convergence reasons \cite{Flensberg}).
Using contour integration to evaluate the Matsubara sum over frequencies
we obtain
\begin{equation}
\sigma_{AB}^{(2)}=\frac{1}{Z}\sum_{n,m}\left(E_{m}-E_{n}\right)^{2}e^{-E_{m}}A_{nm}B_{mn}.
\end{equation}
Comparing this result with the Lehmann representation for $\left\langle B\left[H,\left[H,A\right]\right]\right\rangle $
and $\left\langle \left[H,B\right]\left[H,A\right]\right\rangle $
we obtain the important result 
\begin{equation}
\sigma_{AB}^{(2)}=\left\langle B\pd_{\tau}^{2}A\right\rangle =-\left\langle \pd_{\tau}B\pd_{\tau}A\right\rangle .\label{eq:Sigma2}
\end{equation}
Using \eqref{eq:Sigma2} with $A=B=x_{1}$ and $A=B=x_{2}$ together
with the equations of motion \eqref{eq:EoM2nd_x1} and \eqref{eq:EoM2nd_x2}
we obtain
\begin{align}
m_{1}\sigma_{11}^{(2)} & =k_{1}\left\langle x_{1}^{2}\right\rangle +g\left\langle x_{1}^{2}x_{2}\right\rangle +\frac{w}{2}\left\langle x_{1}^{4}\right\rangle \\
m_{2}\sigma_{22}^{(2)} & =k_{2}\left\langle x_{2}^{2}\right\rangle +\frac{g}{2}\left\langle x_{1}^{2}x_{2}\right\rangle 
\end{align}
so that the interaction energies can be expressed as
\begin{align}
W_{3} & =m_{2}\sigma_{22}^{(2)}-2V_{2}\\
W_{4} & =\frac{1}{4}\left(m_{1}\sigma_{11}^{(2)}-2m_{2}\sigma_{22}^{(2)}\right)-\frac{1}{2}V_{1}+V_{2}.
\end{align}
The kinetic energy terms can also be expressed in terms of $\sigma_{11}^{(2)}$
and $\sigma_{22}^{(2)}$ using \eqref{eq:Sigma2} together with \eqref{eq:x_p_1}
and \eqref{eq:x_p_2}
\begin{equation}
T_{a}=\frac{1}{2}m_{a}\sigma_{aa}^{(2)},
\end{equation}
\label{eq:MGK_simple}and the potential energies are given by
\begin{equation}
V_{a}=\frac{1}{2}k_{a}\frac{1}{\beta}\sum_{i\omega_{n}}G_{aa}(i\omega_{n}).
\end{equation}
Putting all the pieces together, the total energy is given by
\begin{align}
\left\langle H\right\rangle  & =\frac{1}{4\beta}\sum_{i\omega_{n}}\left(3m_{1}\left(i\omega_{n}\right)^{2}+k_{1}\right)G_{11}(i\omega_{n})\nonumber \\
 & +\frac{1}{\beta}\sum_{i\omega_{n}}m_{2}\left(i\omega_{n}\right)^{2}G_{22}(i\omega_{n}).\label{eq:MGK_toy}
\end{align}
Applying Eq.~\eqref{eq:MGK_toy} for the crystalline membrane Hamiltonian
\eqref{eq:Hamiltonian}, we obtain Eqs.~\eqref{eq:U_out} and \eqref{eq:U_in}
of the main text.

%\bibliography{BibliogrGrafeno5}


\end{document}